\documentclass[twocolumn]{aastex62}
\usepackage{amsmath,amstext,eqnarray}
\usepackage[T1]{fontenc}
\usepackage{apjfonts} 
\usepackage[figure,figure*]{hypcap}
\usepackage{commath}
\usepackage{longtable}
\usepackage{CJKutf8}
\usepackage[flushleft]{threeparttable}

\graphicspath{{./}{Figures/}}

\usepackage[none]{hyphenat}
\usepackage{array}
\newcolumntype{+}{>{\global\let\currentrowstyle\relax}}
\newcolumntype{^}{>{\currentrowstyle}}

\shorttitle{Multi-frequency dust polarization of NGC1333 IRAS4A}
\shortauthors{Ko et al.}

\begin{document}

\title{Resolving linear polarization due to emission and extinction of aligned dust grains on NGC1333 IRAS4A with JVLA and ALMA}

\author[0000-0003-2158-8141]{Chia-Lin Ko
\begin{CJK}{UTF8}{bsmi}(柯嘉琳)
\end{CJK}}
\affil{Institute of Astronomy and Department of Physics, National Tsing Hua University, Hsinchu 30013, Taiwan}
\affil{
Academia Sinica Institute of Astronomy and Astrophysics, P.O. Box 23-141, Taipei 10617, Taiwan
}

\author[0000-0003-2300-2626]{Hauyu Baobab Liu
\begin{CJK}{UTF8}{bsmi}(呂浩宇)
\end{CJK}}
\affil{
Academia Sinica Institute of Astronomy and Astrophysics, P.O. Box 23-141, Taipei 10617, Taiwan
}
\affiliation{European Southern Observatory (ESO), Karl-Schwarzschild-Str. 2, 85748, Garching, Germany}

\author[0000-0001-5522-486X]{Shih-Ping Lai
\begin{CJK}{UTF8}{bsmi}(賴詩萍)
\end{CJK}}
\affil{Institute of Astronomy and Department of Physics, National Tsing Hua University, Hsinchu 30013, Taiwan}

\author[0000-0001-8516-2532]{Tao-Chung Ching
\begin{CJK}{UTF8}{bsmi}(慶道沖)
\end{CJK}}
\affil{National Astronomical Observatories, Chinese Academy of Sciences, Beijing 100012, People's Republic of China}
\affil{CAS Key Laboratory of FAST, National Astronomical Observatories, Chinese Academy of Sciences, People's Republic of China}

\author[0000-0002-1407-7944]{Ramprasad Rao}
\affil{
Academia Sinica Institute of Astronomy and Astrophysics, P.O. Box 23-141, Taipei 10617, Taiwan
}

\author[0000-0002-3829-5591]{Josep Miquel Girart}
\affil{Institut de Ci\`encies de l'Espai (ICE, CSIC), Can Magrans, s/n, 08193, Cerdanyola del Vall\`es, Catalonia, Spain}
\affil{Institut d'Estudis Espacials de Catalunya (IEEC), 08034, Barcelona, Catalonia, Spain}

\begin{abstract}

We report high angular resolution
observations of linearly polarized dust emission towards the Class 0 young stellar object (YSO) NGC1333 IRAS4A (hereafter, IRAS4A) using the Karl G. Jansky Very Large Array (JVLA) at K (11.5--16.7 mm), Ka (8.1--10.3 mm), and Q bands (6.3--7.9 mm), and using the Atacama Large Millimeter Array (ALMA) at Band 6 (1.2 mm) and Band 7 (0.85--0.89 mm). 
On 100-1000 AU scales, all of these observations consistently trace the hourglass shaped magnetic field topology as shown in the previous studies. 
In the innermost 100 AU region of IRAS4A1, the polarization position angles (E-field) detected at 6.3--16.7 mm are consistent, however, are nearly 90 degrees offset from those detected at 1.2 mm and 0.85--0.89 mm.  
Such a 90 degree offset may be explained by that the inner $\sim$100 AU area is optically thick at wavelengths shorter than $\sim$1.5 mm, whereby the observations probe the absorption of aligned dust against the weakly or unpolarized warm dust emission from the innermost region.
This can also consistently explain why the highest angular resolution ALMA images at Band 7 show that the polarization percentage increases with dust brightness temperature in the inner $\sim$100 AU region of IRAS4A1.
Following this interpretation and assuming that the dust grains are aligned with the magnetic fields, the inferred magnetic field position angle based on the 90$^{\circ}$ rotated at 6.3--7.9 mm in the central peak of IRAS4A1 is $\sim-22^{\circ}$, which is approximately consistent with the outflow direction $\sim-9^{\circ}$.

\end{abstract}


\section{Introduction}\label{sec:introduction}

Magnetic fields are believed to play a crucial role in the star formation process \citep{Mouschovias1977, Shu1987, Crutcher2012}.
Theoretical studies have suggested that magnetic fields can influence the formation of protostellar disks \citep{Li2014}, and can launch outflows and jets from young protostars \citep{Frank2014}.
Therefore, detailed measurements of magnetic fields are critical to advancing our understanding of star and disk formation in the early stage.

One of the standard method of probing the magnetic field structures is to observe the linearly polarized thermal emission from magnetically aligned elongated dust grains.
When dust grains are aligned with the magnetic field, the observed polarization position angles of dust thermal emission are expected to be perpendicular to the magnetic field lines.
Absorption of continuum background emission by aligned grains can also lead to polarization position angles which are parallel to the magnetic field lines \citep{Hildebrand2000}.
Recent studies also show that self-scattering of thermal dust emission can also produce the polarization seen at (sub)millimeter or centimeter \citep{Kataoka2015, Yang2016}.
The polarization percentage due to dust scattering is maximized when the largest grain size $a_{max} \sim \lambda / 2\pi$, where $\lambda$ is the observing wavelength \citep{Kataoka2015}.
Therefore, multi-wavelength dust polarization observations are a necessary tool to determine the mechanism that cause linearly dust polarization.

NGC1333\,IRAS4A (hereafter IRAS4A) is one of the best-studied, nearby Class 0 young stellar object (YSO) binary \citep[$d\sim$293 pc;][]{Ortiz2018, Zucker2018} in the Perseus Molecular Cloud.
It has at least two components IRAS4A1 and IRAS4A2 separated by 1\farcs
8 (527 AU); they both emanate bipolar outflows \citep{Santangelo2015,Ching2016}.
Previous observations of dust polarization \citep{Akeson1996,Akeson1997,Girart2006,Goncalves2008,Frau2011,Hull2014,Cox2015,Liu2016,Galametz2018} and CO linear polarization \citep{Girart1999,Ching2016} found that on 100-1000 AU scales the magnetic field morphology is consistent with an hourglass-shape while there may be discrepancies on smaller spatial scales.

To further understand the polarization mechanisms of dust on various sizescales and wavelengths, we observed NGC1333\,IRAS4A using Karl G. Jansky Very Large Array (JVLA) at K (11.5--16.7 mm; 18-26 GHz), Ka (8.1--10 mm; 29-37 GHz), and Q bands (6.3--7.5 mm; 40-48 GHz), and using the Atacama Large Millimeter Array (ALMA) at Band 6 (1.3 mm; 234 GHz) and Band 7 (0.87 mm; 345 GHz). 
The details of our multi-wavelength observations and data reduction techniques are described in \autoref{sec:observation}. 
The polarization line segments and polarization percentage are presented in \autoref{sec:result}.
Discussion and analysis of our observations are provided in \autoref{sec:discussions}.
The conclusions are given in \autoref{sec:conclusions}.

\begin{table*}[tpb]
    \caption{NGC1333\,IRAS4A Observational Parameters}
    \label{tab:IRAS4A_obs}
    \vspace{0.1cm}
    \begin{threeparttable}
    \centering
    \makebox[\linewidth]
    {\scriptsize
    \hspace{-1.4cm}
    \begin{tabular}{lcccccccc}
        \hline
        \hline \noalign {\vspace{0.05cm}}
        \textbf{Parameters} & \multicolumn{2}{c}{\textbf{VLA K band}} & \multicolumn{2}{c}{\textbf{VLA Ka band}} & \multicolumn{2}{c}{\textbf{VLA Q band}} & \textbf{ALMA Band 6} & \textbf{ALMA Band 7} {\vspace{0.05cm}} \\
        \hline \noalign {\vspace{0.05cm}}
        Frequency (GHz) & \multicolumn{2}{c}{18-26} & \multicolumn{2}{c}{29-37} & \multicolumn{2}{c}{40-48} & 233-235 & 336-340, 348-352 \\
        Wavelength (mm) & \multicolumn{2}{c}{11.5-16.7} & \multicolumn{2}{c}{8.1-10.3} & \multicolumn{2}{c}{6.3-7.5} & 1.28 & 0.85-0.86, 0.88-0.89 \\
        Field of view ($\arcsec$) & \multicolumn{2}{c}{127} & \multicolumn{2}{c}{85} & \multicolumn{2}{c}{62} & 28 & 17 \\
        Beam size\,\tnote{a} \, ($\arcsec$), PA ($^\circ$) & \multicolumn{2}{c}{0.36$\times$0.32 ($-$77$^\circ$)} & \multicolumn{2}{c}{0.44$\times$0.39 (66$^\circ$)} & \multicolumn{2}{c}{0.71$\times$0.65 ($-$87$^\circ$)} & 0.49$\times$0.31 ($-$7$^\circ$) & 0.25$\times$0.18 ($-$4$^\circ$){\vspace{0.05cm}} \\
        $\sigma_{I}$\,\tnote{b} \, (mJy beam$^{-1}$)  & \multicolumn{2}{c}{0.010} & \multicolumn{2}{c}{0.040} & \multicolumn{2}{c}{0.030} & 2.0 & 9.0 \\
        $\sigma_{Q}$\,\tnote{b} \, (mJy beam$^{-1}$)  & \multicolumn{2}{c}{0.0074} & \multicolumn{2}{c}{0.017} & \multicolumn{2}{c}{0.014} & 0.12 & 0.80 \\
        $\sigma_{PI}$\,\tnote{b} \, (mJy beam$^{-1}$)  & \multicolumn{2}{c}{0.0074} & \multicolumn{2}{c}{0.017} & \multicolumn{2}{c}{0.014} & 0.12 & 0.80 \\
        \hline \noalign {\vspace{0.05cm}}
        Observing date & 2016 Jun. 04 & 2016 Aug. 06 & 2016 Sep. 04 & 2016 Sep. 05 & 2014 Oct. 13 & 2015 Dec. 21 & 2016 Nov. 04 & 2016 Sep. 06 \\
        Array configuration & B & B & B & B & C & D & C40-5 & C36-6 \\
        Available antennae & 26 & 27 & 24 & 26 & 26 & 24 & 43 & 39 \\
        On-source time (mins) & 55 & 57 & 51 & 35 & 57 & 56 & 96 & 107 \\
        Project baseline lengths (meter) & 178-10130 & 216-10650 & 174-5970 & 230-8420 & 36-3140 & 36-986 & 12-1110 & 12-2530 \\
        Project baseline lengths ($k\lambda$) & 13-922 & 10-877 & 169-741 & 22-1040 & 5-460 & 10-877 & 9-865 & 14-2910 \\
        Gain calibrators & J0336+3218 & J0336+3218 & J0336+3218 & J0336+3218 & J0336+3218 & J0336+3218 & J0336+3218 & J0336+3218 \\
        Absolute flux calibrators & 3C147 & 3C147 & 3C147 & 3C147 & 3C147 & 3C147 & J0238+1636, J0336+3218 & J0238+1636\\
        Bandpass calibrators & 3C84 & 3C84 & 3C84 & 3C84 & 3C84 & 3C84 & J0237+2848, J0238+1636 & J0237+2848 \\
        Polarization position angle calibrators & 3C147 & 3C147 & 3C147 & 3C147 & 3C147 & 3C147 & J0334-4008 & J0334-4008\\
        Leakage calibrators & 3C84 & 3C84 & 3C84 & 3C84 & 3C84 & 3C84 & J0334-4008 & J0334-4008 \\
                & J2355+4950 & J0713+4349 & J0713+4349 & J0713+4349 & J0713+4349 & J0713+4349 &  &   {\vspace{0.05cm}}\\
        \hline \noalign {\vspace{0.05cm}}
    \end{tabular}
    }
    \begin{tablenotes}
        \footnotesize
        \item \textbf{Notes.}
        \item[a] Measured from the images of limited $uv$ distance range of 22-865 $k\lambda$ and nature weighted.
        \item[b] Measured from the images of limited $uv$ distance range of 22-865 $k\lambda$, nature weighted and smoothed to 0.$\arcsec$72 resolution.
    \end{tablenotes}
    \end{threeparttable}
\end{table*}

\section{Observations and Data reduction}\label{sec:observation}

\subsection{JVLA Observations}\label{sub:jvla}
We have performed full Stokes polarization observations towards IRAS4A using the JVLA at K band and Ka band in the B array configuration, and at Q band in the C and D array configuration.
The pointing center for our target source was R.A.=03$^{\mbox{\scriptsize{h}}}$29$^{\mbox{\scriptsize{m}}}$10.550$^{\mbox{\scriptsize{s}}}$ (J2000), Decl.=+31$^{\circ}$13$'$31$''$.0 (J2000).
The C array configuration observations at Q band have been introduced in \citet{Liu2016}.
The D array configuration observations at Q band were carried out on 2015 December 21 (project code: 15B-049, PI: Hauyu Baobab Liu).
The B array configuration observations at K and Ka band were carried out from mid to late 2016 (project code: 16A-109, PI: Hauyu Baobab Liu).
All these observations used the 3 bit sampler, and configured the backend to have an 8 GHz bandwidth coverage by 64 consecutive spectral windows.
Other details of these observations are summarized in \autoref{tab:IRAS4A_obs}.

We manually followed the standard data calibration strategy using the Common Astronomy Software Applications \citep[CASA;][]{McMullin2007} package release 5.3.0.
After implementing antenna position corrections, weather information, gain-elevation curve and opacity model, we bootstrapped delay fitting and passband calibrations, and then performed complex gain calibration, cross-hand delay fitting, polarization leakage calibration, and polarization position angle referencing.
We applied the absolute flux reference to our complex gain solutions, and then applied all derived solution tables to the target source.
Finally, we performed 3 iterations of gain phase self-calibrations for the D array configuration observations at Q band, to remove the residual phase offsets.
Our target source is not bright enough at Ka and K band for self-calibrating.

\subsection{ALMA Observations}\label{sub:alma}

We performed polarizaton observations towards IRAS4A at Band 6 (1.28 mm, 234 GHz) and Band 7 (0.87 mm, 344 GHz) using ALMA on 2016 November 4 (2016.1.01089.S, PI: Tao-Chung Ching) and on 2016 September 6 (2015.1.00546.S, PI: Shih-Ping Lai), respectively. 
The ALMA configurations for the 1.28 mm and 0.87
mm observations are C40-5 with the $uv$ distance range from 9 to 865 k$\lambda$, and C36-6 with the $uv$ distance range from 14 to 2910 k$\lambda$, respectively.
We utilized the $uv$ data calibrated by the ALMA Regional Center (ARC) using the CASA version 4.6.0.
Other details of these observations are summarized in \autoref{tab:IRAS4A_obs}.

\subsection{Polarization Images}\label{sub:pol}

We imaged the calibrated data using the \texttt{tclean} task of the CASA version 5.3.0.
We produced images of polarization intensity ($PI$), polarization percentage (P), and polarization position angle ($PA$) from Stoke $I$, $Q$, $U$ maps using the CASA task \texttt{immath}.
The polarization intensity is debiased using the formula $PI$ = $\sqrt{Q^2 + U^2 - \sigma_{PI}^2}$, where $\sigma_{PI}$ is assumed to be the average noise level determined from the stoke $Q$ and $U$ maps for simplicity \citep{Simmons1985,Vaillancourt2006}.
We evaluated $PA$ as 0.5 $\arctan$\,($U$/$Q$) and $P$ as ($PI/I$)$\times$100\% from where the $I$ and $PI$ intensities were both detected at $>$3 $\sigma$ significance.
We follow a convention that $PA$ is defined in between $\pm$ 90$^{\circ}$; north and east are $PA$ = 0$^{\circ}$ and 90$^{\circ}$, respectively.

\section{Results}\label{sec:result}

\autoref{fig:pol_PI_E}(a) and \autoref{fig:pol_PI_E}(c) present polarization line segments taken from all observations overlaid on the 6.9 mm brightness temperature ($T_{B}^{\mbox{\scriptsize 6.9 mm}}$) and the 6.9 mm to 0.87 mm brightness temperature ratio ($T_{B}^{\mbox{\scriptsize 6.9 mm}}$/$T_{B}^{\mbox{\scriptsize 0.87 mm}}$), respectively.
The images were produced using natural weighting (i.e., Briggs weighting with a robust parameter of 2) to yield the best possible signal-to-noise ratio.
In order to compare the same scales for all the different wavelengths, we used visibilities in the $uv$ range of 22--865 $k\lambda$ and smoothed all images to 0\farcs72 resolution.
Our measurements of polarization position angle, percentage, intensity, and Stokes $I$, $Q$ and $U$ and their uncertainties at all wavelengths are listed in \autoref{tab:IRAS4A_value} in \autoref{sec:pol_measurements}.
The polarization maps at the five observed wavelength bands generated using natural weighting and the limited $uv$ distance range of 22--865 $k\lambda$, are provided in  \autoref{fig:pol_PI_E_all} in \autoref{sec:pol_images}.
At projected radii greater than 150 AU from IRAS4A1, the polarization position angles at all bands are consistent with each other, and trace the hourglass-shaped magnetic field shown by previous studies \citep[e.g.][]{Girart2006, Hull2014}.  
In the central 150 AU around IRAS4A1, the polarization position angles detected with JVLA are nearly 90$^\circ$ offset from  those detected with ALMA.

A zoom-in of \autoref{fig:pol_PI_E}(a) towards IRAS4A1 is presented in \autoref{fig:pol_PI_E}(b). 
To obtain the detailed distribution at higher angular resolution of polarization line segments around IRAS4A1, we made additional images using the same visibility range but with a robust = $-$1 for the JVLA Q band data and robust = $-$2 for the ALMA Band 6 data. 
The resulting images were smoothed to a final 0\farcs45 angular resolution. 
The 90$^\circ$ offsets are only seen in the innermost 100 AU around IRAS4A1.
These approximately 90$^\circ$ offsets are also displayed in
\autoref{fig:PA_sub_BT_Q}(a) and \ref{fig:PA_sub_BT_Q}(b), which show the difference in polarization position angles between ALMA Band 7 and the other bands.

\autoref{fig:PA_sub_BT_Q}(c) shows the polarization percentages in the innermost 100\,AU region of IRAS4A1, which were measured from images of the five observed wavelength bands generated with 0\farcs72 angular resolution.
We caution that the measurements of polarization percentage may be biased by the beam smearing.

\autoref{fig:pol_PI_E}(d) displays the polarization percentage at ALMA Band 7 overlaid with the polarization line segments using nature weighting.
The polarization percentage is generally higher (> 10\%) in the outer region, dropping to less than 1\% towards both IRAS4A1 and IRAS4A2.
However, in the innermost 100 AU around IRAS4A1, the polarization percentage stops decreasing and climbs back to $\sim$4\%.
\autoref{fig:PA_sub_BT_Q}(d) shows the observed polarization percentage versus brightness temperature at 0.87\,mm.
In addition, in \autoref{fig:PA_sub_BT_Q}(e) and \ref{fig:PA_sub_BT_Q}(f) we plot such measurements from the areas which are enclosed by the lowest 6.9\,mm stoke I iso-intensity contours (1.1 mJy beam$^{-1}$) that isolate IRAS4A1 and IRAS4A2. 
For IRAS4A1, the polarization percentage decreases at $T_{B}^{\mbox{\scriptsize 0.87 mm}}<30K$ and increases at $T_{B}^{\mbox{\scriptsize 0.87 mm}}>30K$, while for IRAS4A2, the polarization percentage decreases as $T_{B}^{\mbox{\scriptsize 0.87 mm}}$ increases.

\section{Discussion}\label{sec:discussions}

The observed dust linear polarization can be attributed to aligned dust grains.
In the envelope region ($\sim$100--1000\,AU), the brightness temperature ratio $T_{B}^{\mbox{\scriptsize 6.9 mm}}$/$T_{B}^{\mbox{\scriptsize 0.87 mm}}$ is low ($\sim$0.1), which indicates that dust is likely optically thin at all observed wavelengths. 
Hence the polarization position angles at all observed wavelengths are parallel to the projected long axis of dust grains.
If the dust grains are aligned perpendicular to the magnetic field lines, then the magnetic field morphology inferred from our observations is consistent with a hourglass shape.
The polarization observations of CO lines are also trace the consistent magnetic field morphology \citep{Girart1999, Ching2016}.

In the innermost $\sim$100\,AU region around IRAS4A1 (i.e., within the central one beam area of 0\farcs72 resolution), our observed brightness temperature ratio $T_{B}^{\mbox{\scriptsize 6.9 mm}}$/$T_{B}^{\mbox{\scriptsize 0.87 mm}}$ is $\sim$0.6. 
We derive the optical depth ($\tau$) at 6.9 mm  by assuming that the brightness temperature of dust emission is $T_{B}$ = $T_{\mbox{\scriptsize dust}} (1 - e^{-\tau})$, where the dust temperature ($T_{\mbox{\scriptsize dust}}$) can be factored out.
The optical depth at 0.87 mm is approximately $\tau^{\mbox{\scriptsize\,0.87mm}}$ = $\tau^{\mbox{\scriptsize\,6.9 mm}}$ $\cdot$ (6.9\,mm\,/\,0.87\,mm)$^{\beta}$, where the dust opacity index $\beta$ is assumed to be 2.0. 
The corresponding value of $\tau^{\mbox{\scriptsize\,6.9 mm}}$ in the innermost $\sim$100 AU region of IRAS4A1 is $\sim$0.92, which suggests that the polarization at or longer than 6.9 mm is optically thin or marginally optically thick \citep{Liu2018}.
Thus, the polarization observation with JVLA is likely contributed by polarized dust emission, which is similar to the mechanism at the outside 100 AU region of IRAS4A1.
On the other hand, the corresponding value of $\tau^{\mbox{\scriptsize\,0.87 mm}}$ is $\sim$58, which indicates that the dust emission at 0.87 mm and 1.2 mm is already optically thick \citep{Liu2016, Li2017, Sahu2019}

The polarization percentages at all wavelengths in the inner $\sim100$ AU of IRAS4A1 are in the range of 1--4 \% and show no obvious relation with wavelength (\autoref{fig:PA_sub_BT_Q}c).
In addition, based on the analysis of Stokes $I$ emission at (sub)millimeter and centimeter wavelength bands, 
\citet{Li2017} found that the dust opacity spectral index is consistent with $\sim$2, which are not distinguishable from that in the diffuse interstellar medium (for a more recent discussion see \citealt{Galvan-Madrid2018}).
Conventionally, the $\sim$2 values of dust opacity spectral index would infer that the maximum grain sizes are less than  100 $\mu$m  (c.f., \citealt{Testi2014} and references therein).
The inferred values of maximum grain sizes may be still smaller after the effects of dust scattering opacity is self-consistently considered (\citealt{Liu2019ApJ}, \citealt{Zhu2019ApJ}, and \citealt{Takaki2019}).
If it is indeed the case, then the polarization due to dust self-scattering should be non-detectable at wavelengths longer than 6.9 mm (c.f., Figure 3 of \citealt{Kataoka2016}).
Recent studies of the Class II YSOs also suggests that the maximum grain size should be $\lesssim$100 $\mu$m \citep[e.g.,][]{Kataoka2017, Hull2018, Dent2019, Okuzumi2019, Stephens2017, Ohashi2018}, inferring that the maximum polarization generated by the self-scattering is $\lesssim$1\,\% \citep{Kataoka2015}.
Given that dust grains in Class 0 YSOs are likely smaller than those in Class II YSOs, we argue that on the spatial scale and wavelength range probed by our polarization observations, dust self-scattering may not be a significant polarization mechanism.
As to the radiative alignment, we can not rule out this possibility without spatially resolving the innermost $\sim$100 AU region of IRAS4A1.
Nevertheless, we disfavor the radiative alignment due to that it predicts no polarization at the central location.

We hypothesize that in this case the JVLA observations are indicative of polarized emission from aligned dust grains, while the ALMA observations point towards extinction due to aligned dust grain.
This can help to explains the 90$^\circ$ offset of the polarization position angles observed by these two instruments.
Following this interpretation, we hypothesize that at wavelengths longer than 6.9 mm, the 90$^{\circ}$-relationed polarization position angles trace the projected B-field orientation (\autoref{fig:B-field}).
The inferred B-field position angles in the central 100 AU around IRAS4A1 is $\sim-22^{\circ}$, which is approximately consistent with the outflow position angles of $\sim-9^{
\circ}$ \citep{Ching2016}.

In \autoref{fig:PA_sub_BT_Q}(d)-(f), we provide a simplified model to qualitatively explain the observed polarization percentage distribution at 345 GHz.
Our fiducial simplified model is composed of two components: one foreground component, which can be illustrated as the circumstellar envelope on $\sim$100--1000 AU scales; and one (foreground) obscured background component, which can be illustrated as a marginally resolved disk on $\lesssim$100 AU scales, obscured by the foreground circumstellar envelope.

Assuming that dust on 100-1000 AU scales is predominantly heated by the stellar irradiation from IRAS4A1, and assuming an approximate thermal equilibrium, the dust temperature $T_{\mbox{\scriptsize dust}}$ should have an $r^{-\frac{1}{2}}$ dependence according to the Stefan-Boltzmann law, where $r$ is the radius from the host protostar of IRAS4A1.
Assuming that the mass volume density around IRAS4A1 can be approximated by a singular isothermal sphere \citep{Shu1977ApJ}, then the volume density $n$ has a $r^{-2}$ dependence.
The dust column density $\Sigma_{\mbox{\scriptsize dust}}$ and optical depth $\tau$ then have  $r^{-1}$ dependence.
Therefore, in the envelope we approximately have $T_{\mbox{\scriptsize dust}}\propto\tau^{0.5}$.

Based on the SED fittings, \citet{Li2017} suggested that the envelope component has a dust temperature $T_{\mbox{\scriptsize dust}}\sim$20 K and the dust optical depths $\sim$1 in the inner $\sim$100 AU region.
Motivated by these measurements, we adopted the $T_{\mbox{\scriptsize dust}}$ values and the dust optical depth at the two orthogonal linear polarization orientation (E and B), $\tau^{E}$ and $\tau^{B}$, following:

\begin{equation}
T_{\mbox{\scriptsize dust}} = 20 [K] \times \tau^{\frac{1}{2}}, \,\,\,\,\,\,
\tau^{B} \equiv (1 - \alpha)\tau, \,\,\,\,\,\,
\tau^{E} \equiv (1 + \alpha)\tau,
\end{equation}
where $\alpha$ is the polarization efficiency.
We assume $\alpha$ to be a 5\% constant.
We evaluated the brightness temperatures $T^{B}$ and $T^{E}$ on the area where the fore/background components are not overlapped by
\begin{equation}
T^{B} = \frac{1}{2}T_{\mbox{\scriptsize dust}} (1 - e^{-\tau^{B}} ), \,\,\,\,\,\,
T^{E} = \frac{1}{2}T_{\mbox{\scriptsize dust}} (1 - e^{-\tau^{E}} ),
\end{equation}
The Stokes I intensity ($I$), polarized intensity ($PI$), and polarization percentage ($P$) are $T^{B} + T^{E}$, $T^{B} - T^{E}$, and $|PI/I|$, respectively.
By varying $\tau$, the obtained $P$ as a function of $I$ is shown as the orange curve in \autoref{fig:PA_sub_BT_Q}(e) and \ref{fig:PA_sub_BT_Q}(f) to compare with the observations around IRAS4A1 and IRAS4A2, respectively.
For IRAS4A2, the observed polarization percentage is below the predicted model as shown in \autoref{fig:PA_sub_BT_Q}(f).
This may partly due to that there are a forest of spectral lines \citep{Sahu2019}, which leads to an overestimate of stoke I continuum flux and consequently an underestimate of polarization percentage.
For comparison, we have also done the similar evaluation but assuming a constant $T_{\mbox{\scriptsize dust}}=$ 20 K, which is shown as the blue curve in Figure \ref{fig:PA_sub_BT_Q}(b).

We assumed that the background component (e.g., the disk) has a $>$45 K dust temperature ($T^{bg}_{\mbox{\scriptsize dust}}$) distribution, which was also motivated by the SED fittings of \citet{Li2017}.
We further assumed that in the area that the foreground component (e.g., the envelope) is obscuring the background component, the foreground component has an average dust temperature $\sim$20 K and the average optical depth $\tau\sim$1.
We evaluated $T^{B}$ and $T^{E}$ by
\begin{align}
T^{B} = & \frac{1}{2}T^{bg}_{\mbox{\scriptsize dust}}e^{-\tau^{B}}+\frac{1}{2}T_{\mbox{\scriptsize dust}}(1-e^{-\tau^{B}}), \\
T^{E} = & \frac{1}{2}T^{bg}_{\mbox{\scriptsize dust}}e^{-\tau^{E}}+\frac{1}{2}T_{\mbox{\scriptsize dust}}(1-e^{-\tau^{E}}),
\end{align}
and then $I$, $PI$, and $P$ was evaluated by varying the values of $T^{bg}_{\mbox{\scriptsize dust}}$ and fixing the other parameters \citep[c.f.,][]{Liu2018}. 
The resulting fitted $P$ as a function of $I$ is shown as the green curve in \autoref{fig:PA_sub_BT_Q}(e), which covers the innermost $\sim$100 AU region of IRAS4A1.

This simple two components model can qualitatively explain the observed trends of $P$ at 345 GHz.
We did not intend to perform detailed fitting due to the uncertainty in the assumption of $\alpha$, the uncertainties of $P$ due to polarization canceling in a finite synthesized beam, and the missing fluxes over extended angular scales which can bias high the observed $P$ at regions with low $I$.
Furthermore our target source is in fact a binary which likely has a more complicated density and temperature structure than what can be reproduced by assuming very few free parameters.

\section{Conclusions}\label{sec:conclusions}

We have performed full polarization observations with JVLA at 11.5--16.7, 8.1--10.3 and 6.3--7.5 mm, and with ALMA at 1.3 and 0.85--0.89 mm towards the Class 0 YSO NGC1333\,IRAS4A.
We have successfully detected linear polarization from all observations.
Our JVLA K-band data offer the dust polarization ever detected at longest wavelengths in YSOs.
We found that the polarization angles from all bands are consistent with each other on the larger 100--1000 AU scales from IRAS4A1, while on the innermost $\sim$100 AU scales around IRAS4A1 the polarization angles of JVLA data and ALMA data are approximately perpendicular to each other.
This 90$^{\circ}$ offset can be explained if the polarization originates from aligned dust grains at different optical depths; 
the dust emission is marginally optically thick or optically thin at JVLA bands and is optically thick at ALMA Bands.
Thus the polarization at ALMA bands suffers from foreground extinction which leads to the observed polarization being perpendicular to the magnetic field.
In addition, the variation of polarization can also be modeled with polarization of aligned dust grains from an optically thick background source with a foreground envelope.
The magnetic field direction around IRAS4A1 inferred from the aligned dust grains is $\sim-22^{\circ}$, which is very close to the outflow position angles of $\sim-9^{\circ}$, suggesting that magnetic fields are important for launching outflows.

\begin{figure*}
   \begin{tabular}{ p{8.5cm} p{8.5cm} }
     \includegraphics[width=9.2cm]{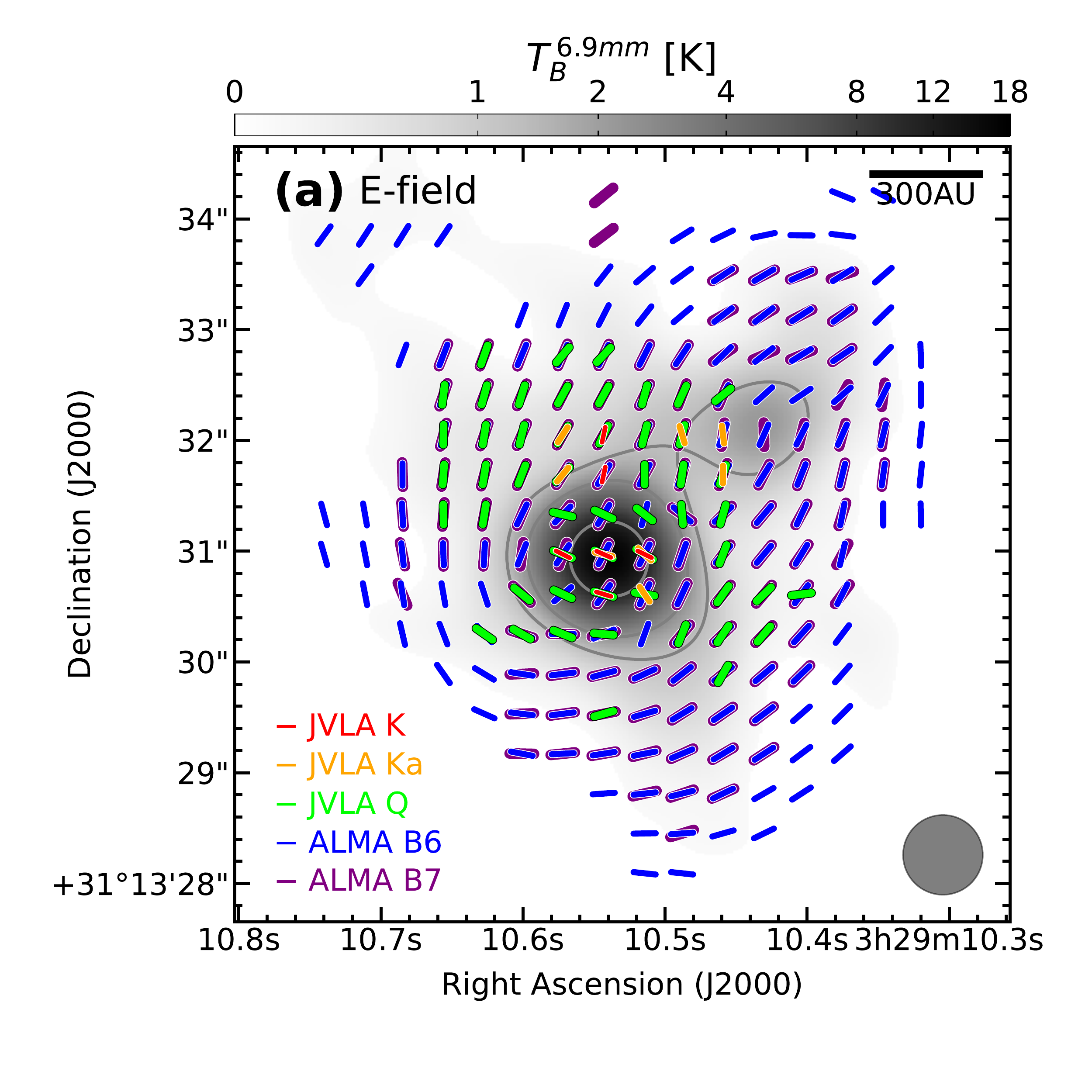} &
	 \includegraphics[width=9.2cm]{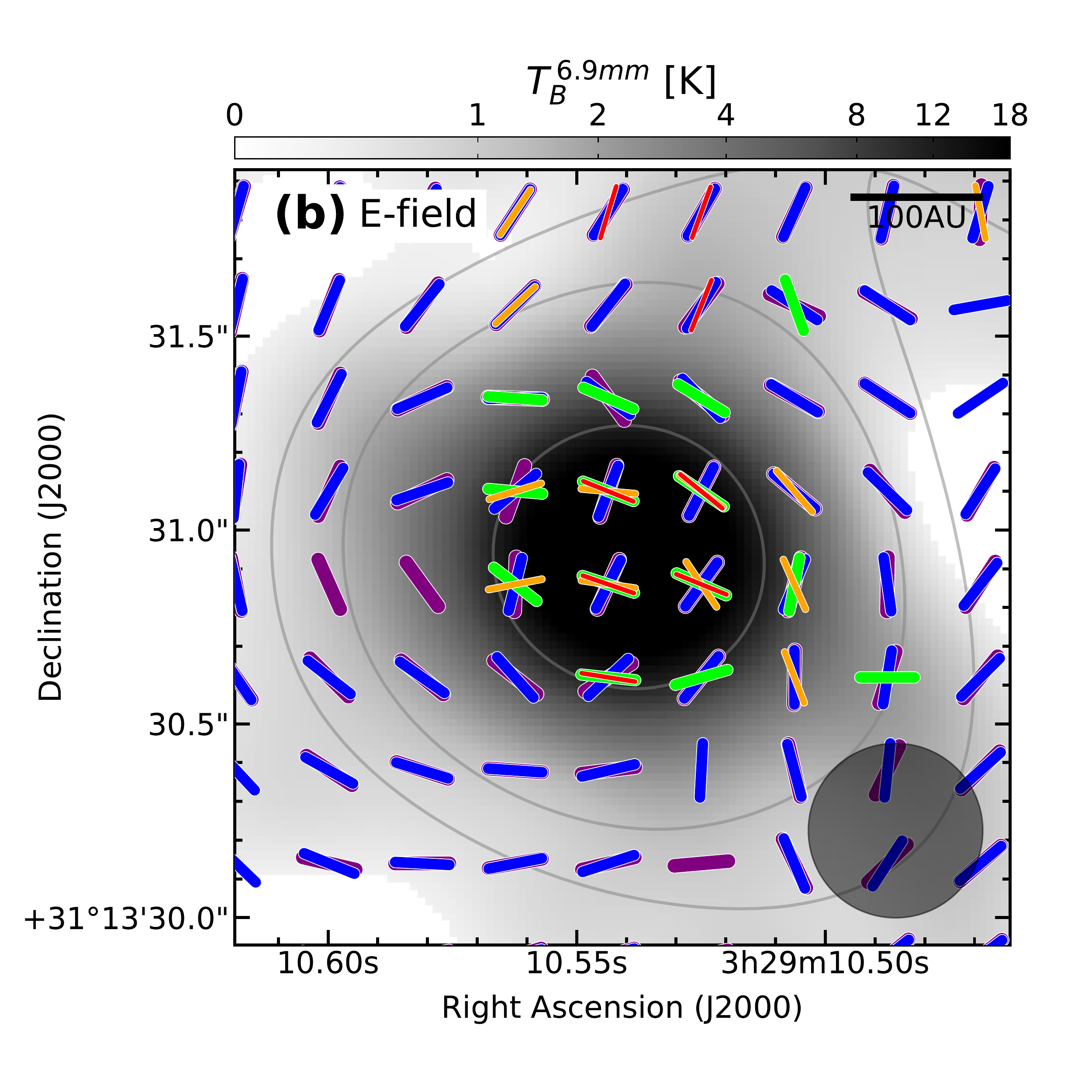}\\
   \end{tabular}
   
   \begin{tabular}{ p{8.5cm} p{8.5cm} }
     \includegraphics[width=9.2cm]{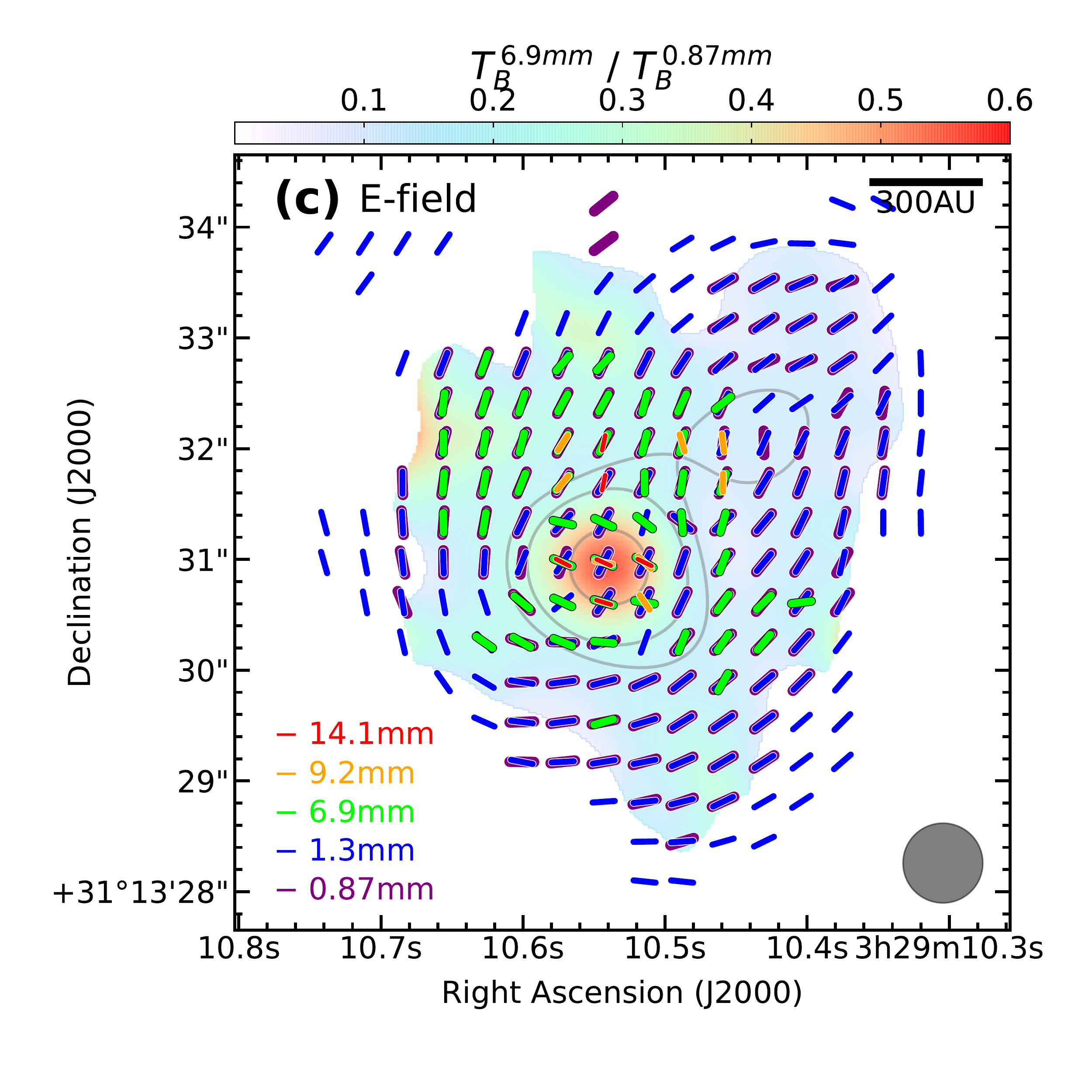} &
     \includegraphics[width=9.2cm]{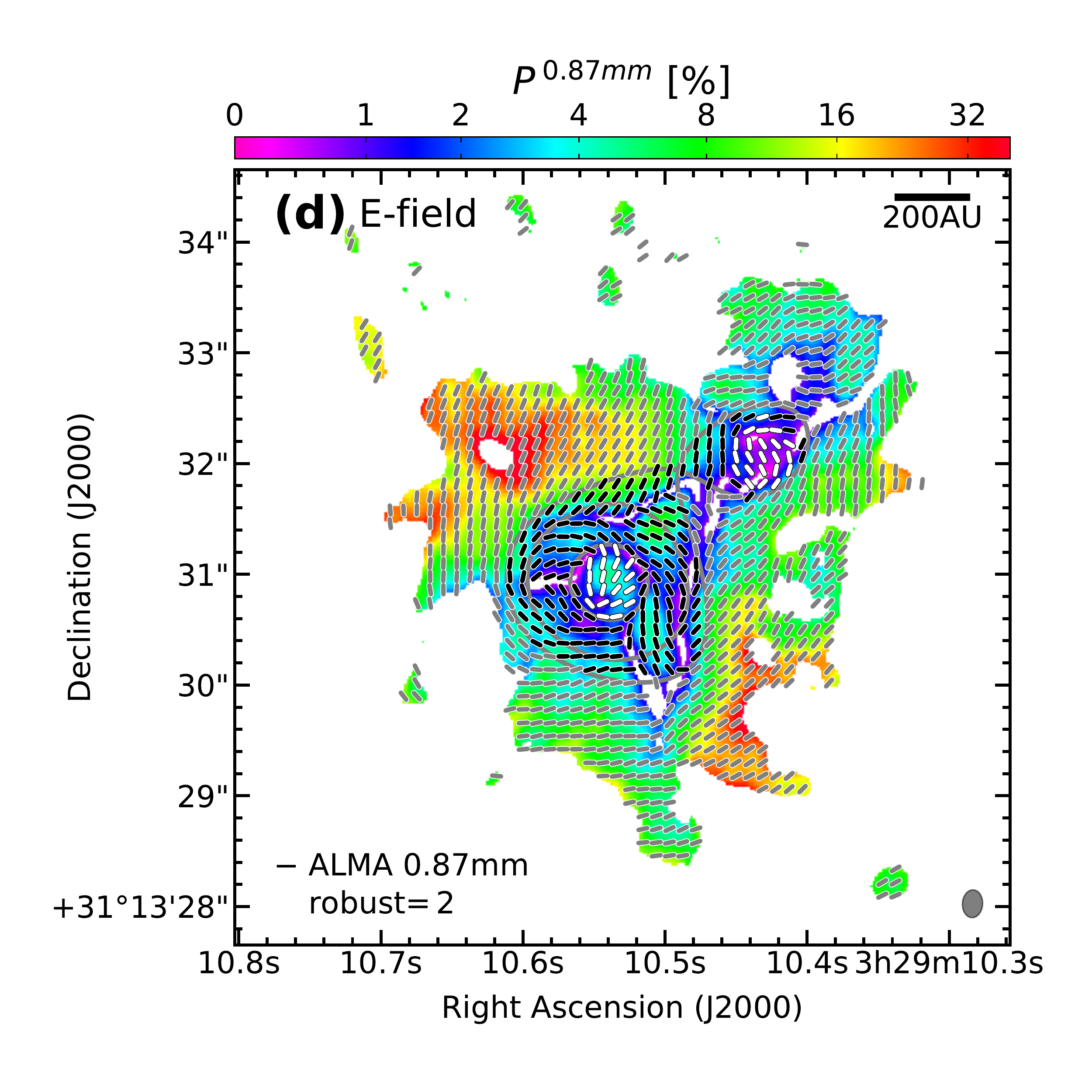}\\
   \end{tabular}
   \caption{\footnotesize{
   JVLA K (14.1 mm), Ka (9.2 mm), and Q bands (6.9 mm), and ALMA Band 6 (1.3 mm) and Band 7 (0.87 mm) full polarization observations on NGC1333\,IRAS4A.
   (a) The brightness temperature in JVLA 6.9 mm (grayscale) overlaid with the polarization orientations (color line segments) obtained from the smoothed to 0\farcs72 resolution images.
   (b) Zoomed image of the panel (a) with the 0\farcs45 resolution.
   (c) The brightness temperature ratio of JVLA 6.9mm to ALMA 0.87mm in colorscale, and the other parameters are the same as the panel (a).
   (d) The polarization percentage in ALMA 0.87 mm (colorscale) overlaid with the polarization angle (PA) with an angular resolution of $\theta_{\mbox{\scriptsize maj}}$ $\times$ $\theta_{\mbox{\scriptsize min}}$ = 0\farcs25 $\times$ 0\farcs18 (P.A. = $-3.6^\circ$).
   The white line segments represent the PA at the innermost 100 AU region of IRAS4A1 and IRAS4A2.
   The black line segments represent the PA at the outside $>$ 35.5 $\times$ $\sigma_{\mbox{\scriptsize I}}$ area of the innermost region of IRAS4A1 and IRAS4A2, where $\sigma_{\mbox{\scriptsize I}}$ = 30 $\mu$Jy\,beam$^{-1}$ for the JVLA 6.9 mm data.
   The gray line segments represent the PA at the remaining region.
   The red, orange, green, blue, and purple line segments present the E-field polarization position angles observed at both $>$3 $\sigma_{\mbox{\scriptsize I}}$ and $>$3 $\sigma_{\mbox{\scriptsize PI}}$ significance for JVLA 14.1 mm, 9.2 mm, and 6.9 mm, and ALMA 1.3 mm and 0.87 mm, respectively.
   All the observations are within a limited $uv$ distance range of 22-865 $k\lambda$.
   The levels of gray contours are [1, 2, 4] $\times$ 1.1 mJy\,beam$^{-1}$
   .
   The synthesized beams are shown in the bottom right of each panel.
   The scale bars are based on the assumption of a 293 pc distance.
   }
   \label{fig:pol_PI_E}
   }
\end{figure*}

\begin{figure*}
   \begin{tabular}{ p{8.8cm} p{8.8cm} }
        \includegraphics[width=9.2cm]{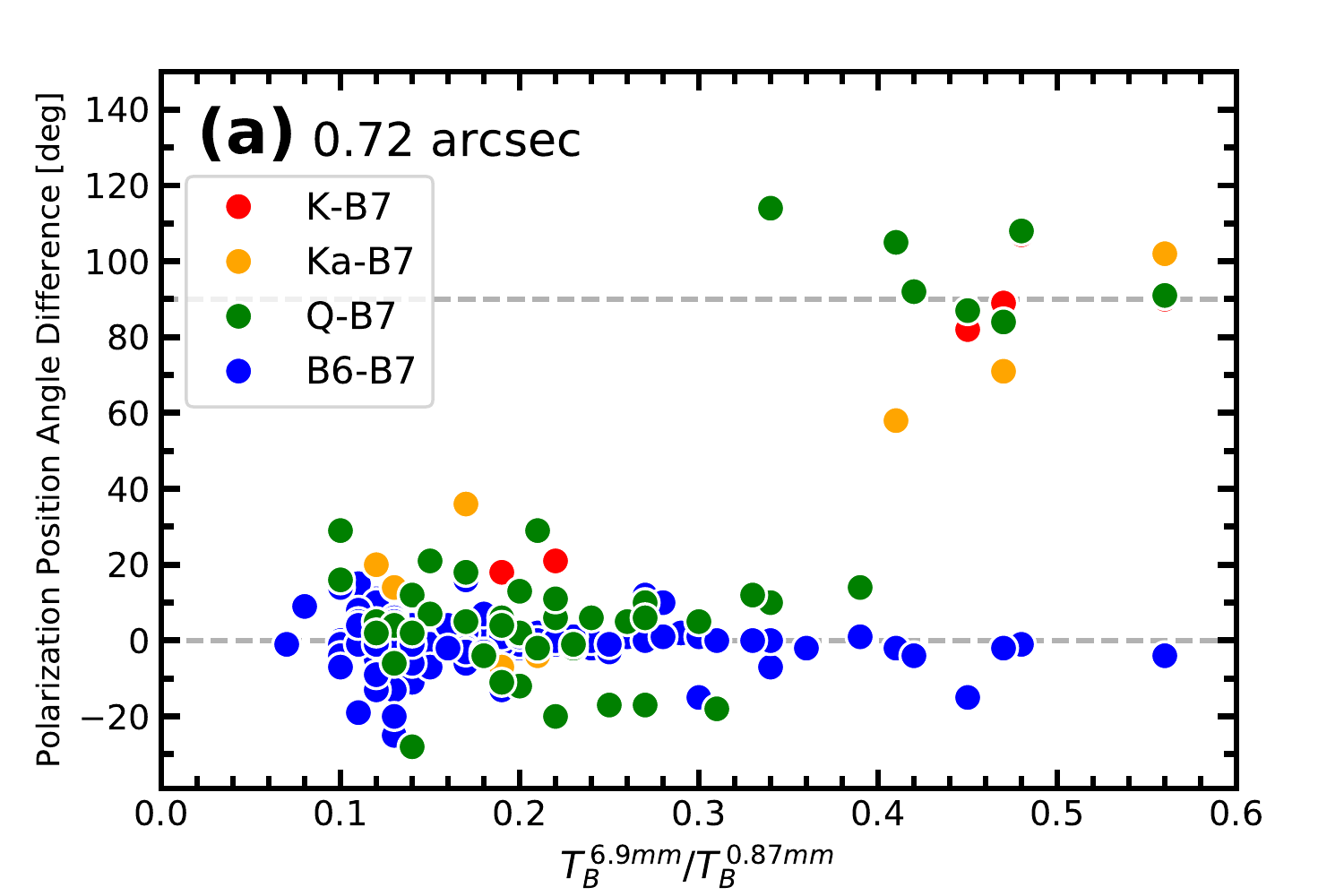} &
        \includegraphics[width=9.2cm]{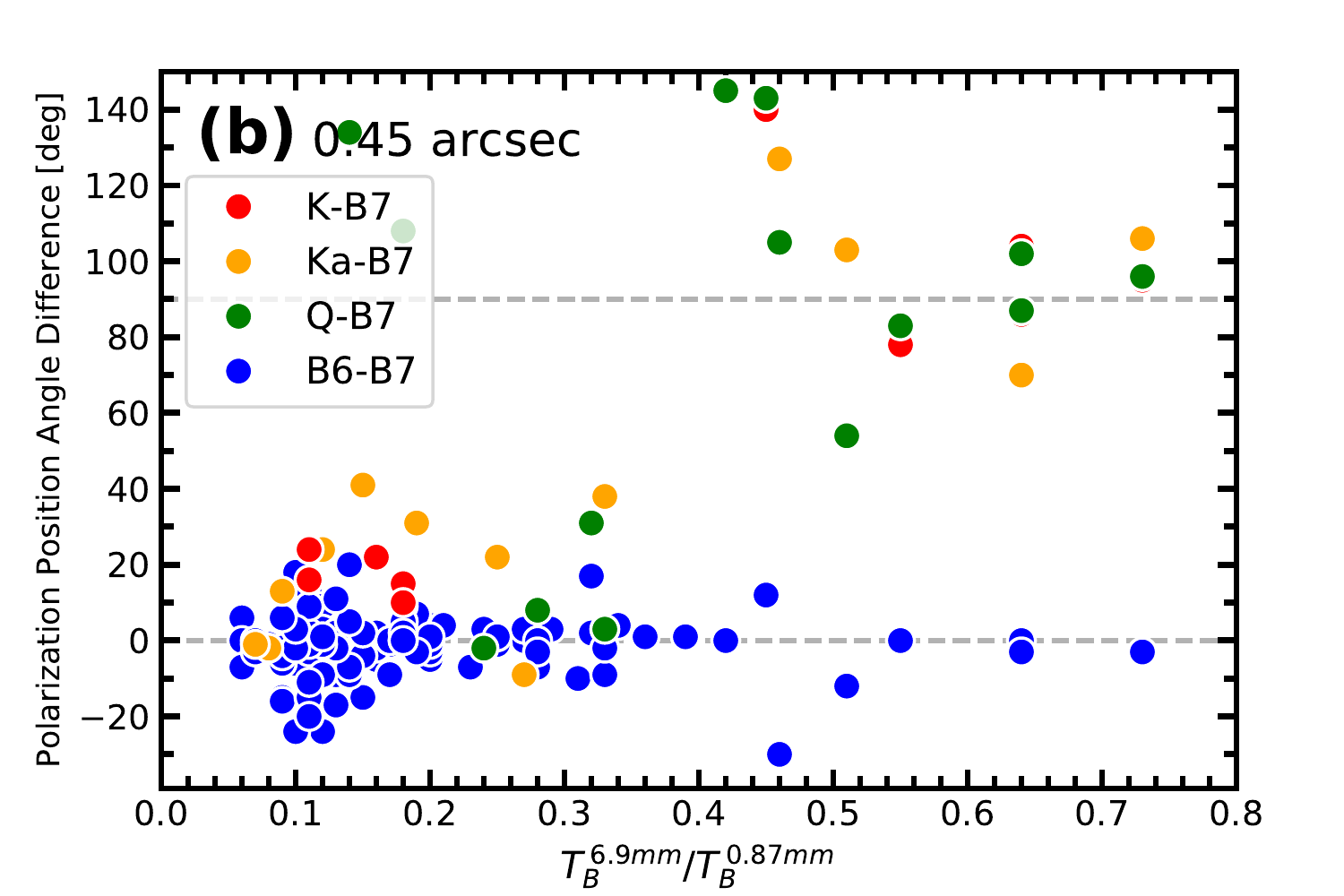}
   \end{tabular}
   
   \begin{tabular}{ p{8.8cm} p{8.8cm} }
        \includegraphics[width=9.2cm]{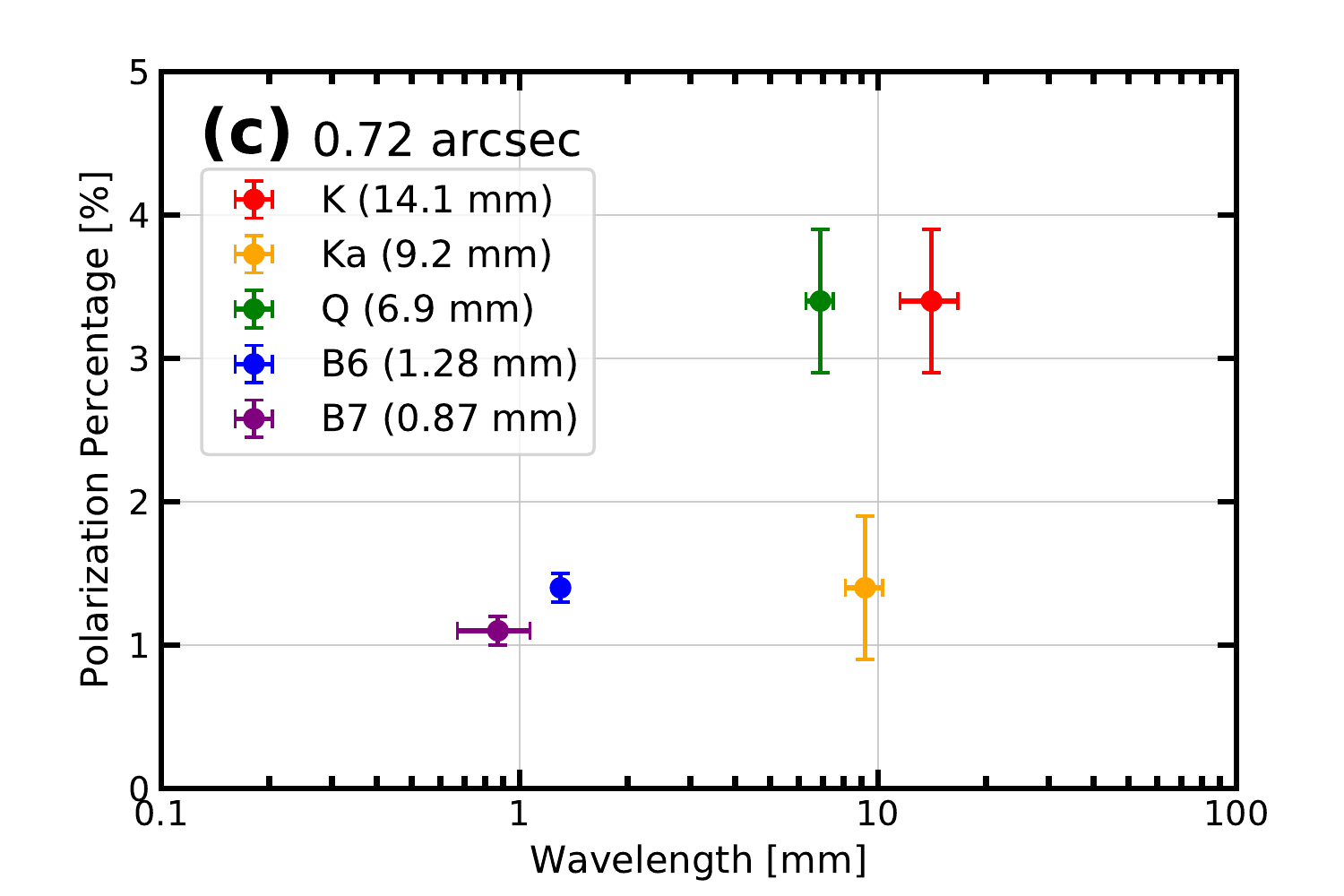} &
        \includegraphics[width=9.2cm]{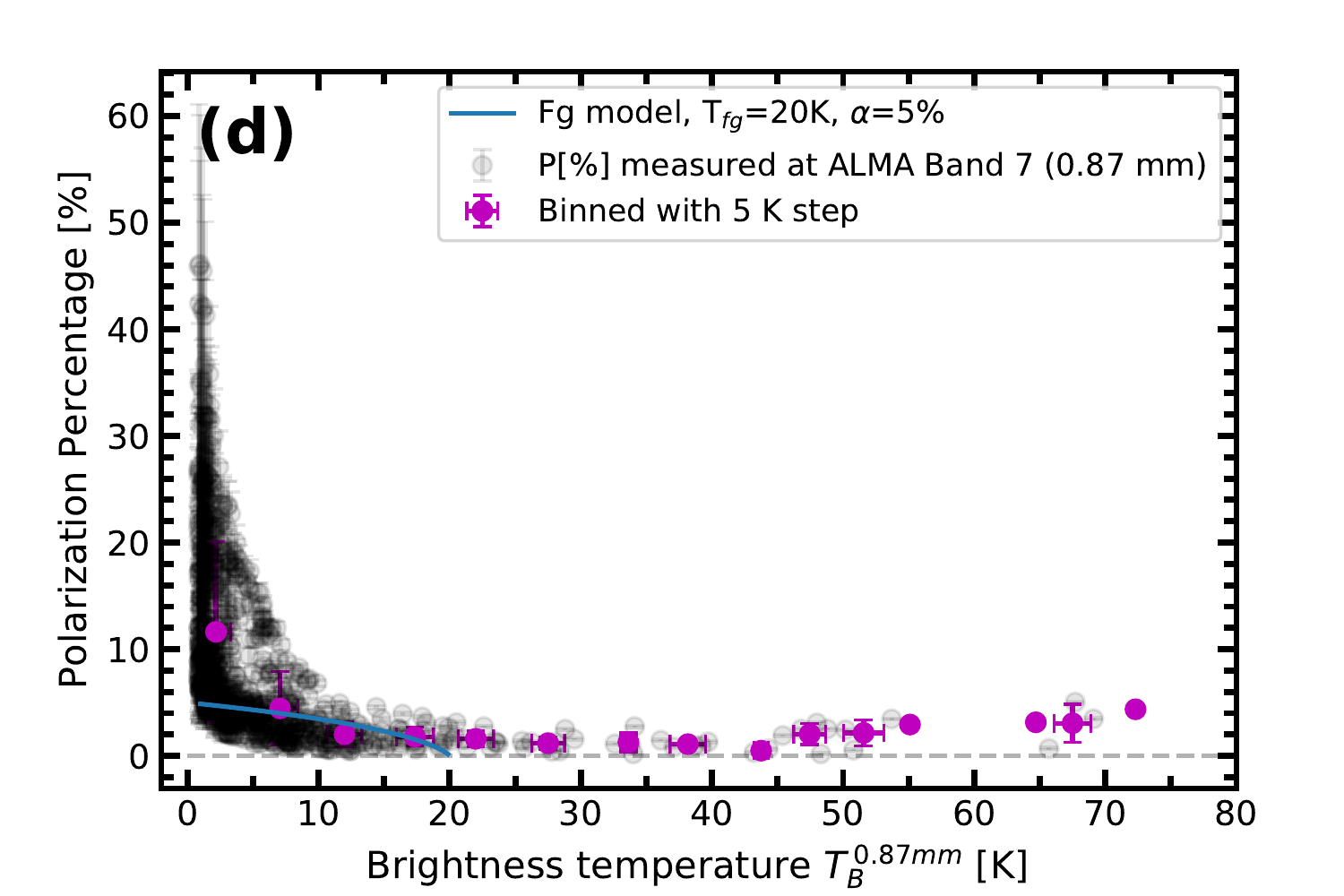}
   \end{tabular}
   
   \begin{tabular}{ p{8.8cm} p{8.8cm} }
     \includegraphics[width=9.2cm]{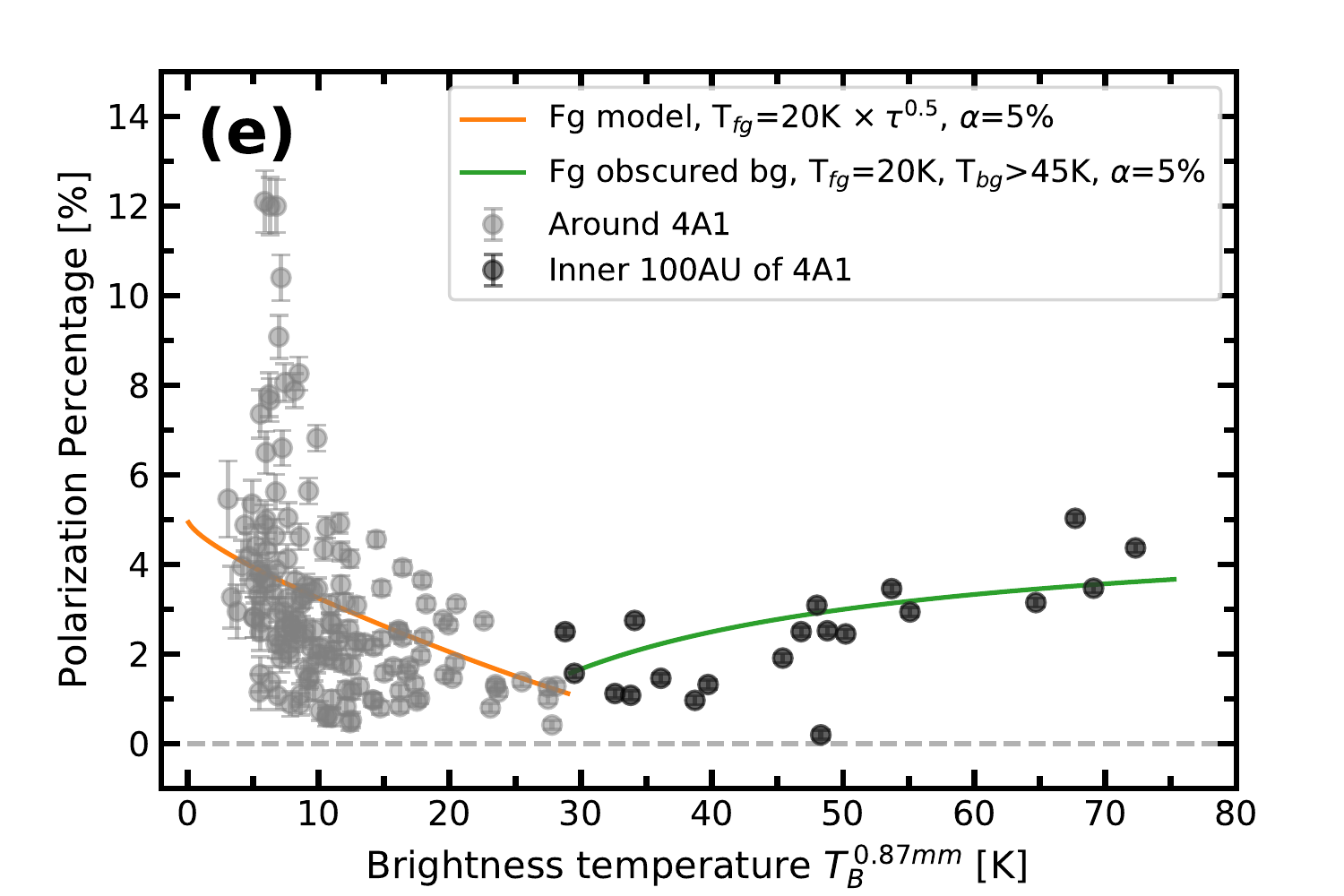} &
     \includegraphics[width=9.2cm]{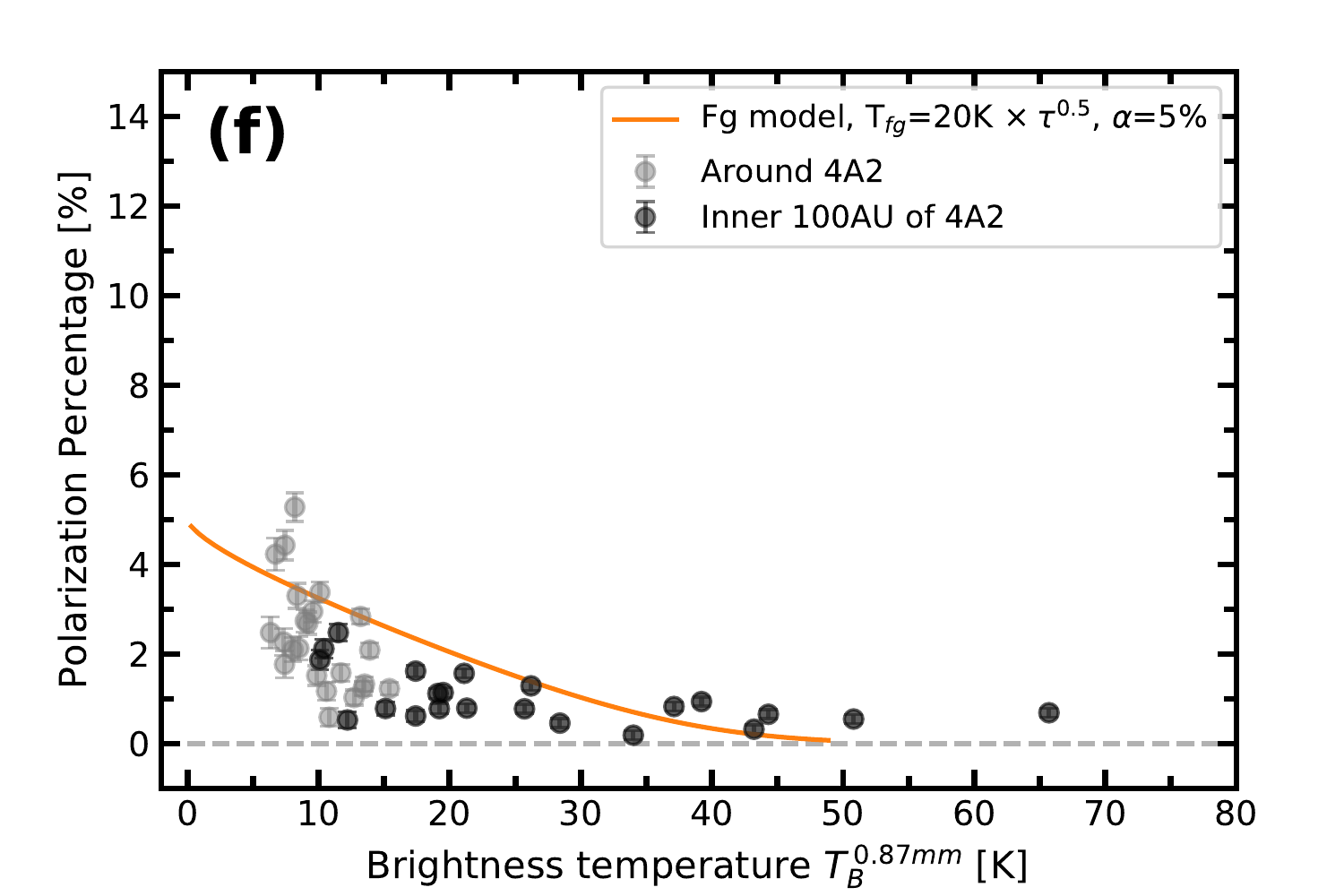}\\
   \end{tabular}
   
   \caption{\footnotesize{
   A summary of polarization positional angle difference and polarization percentage.
   (a) Difference in polarization position angles between ALMA 0.87 mm and JVLA 14.1 mm (K band, red dots), 9.2mm (Ka band, orange dots), 6.9mm (Q bands, green dots) and ALMA 1.3 mm (Band 6, blue dots) plotted against the brightness temperature ratio of JVLA 6.9 mm to ALMA 0.87 mm with the 0\farcs72 resolution.
   The black dash lines present the angles of 0 and 90 degrees.
   (b) Similar to panel (a) but using the images generated with 0\farcs45 resolution.
   (c) Polarization percentages in the innermost 100\,AU region around IRAS4A1 measured from the 0\farcs72 resolution images.
   (d) Relation between the polarization percentage and the brightness temperature of ALMA 0.87 mm with an angular resolution of $\theta_{\mbox{\scriptsize maj}}$ $\times$ $\theta_{\mbox{\scriptsize min}}$ = 0\farcs25 $\times$ 0\farcs18 (P.A. = $-3.6^\circ$).
   The black points show the original values, and the purple points present the values binned with 5 K step.
   The blue line represent the model only a foreground envelope with a constant dust temperature $T_{\mbox{\scriptsize dust}}$ = 20 K.
   (e) Relation between the polarization percentage and the brightness temperature of ALMA 0.87 mm around IRAS4A1.
   The orange line presents the model of one foreground component with $T_{\mbox{\scriptsize dust}} = 20 [K] \times \tau^{0.5}$, where $\tau$ is from 0 to 2.5.
   The green line shows the model of a foreground envelope obscured background marginally resolved disk with $T^{fg}_{\mbox{\scriptsize dust}}$ = 20 K and $T^{bg}_{\mbox{\scriptsize dust}}$ from 45 K to 170 K, respectively.
   (f) Relation between the polarization percentage and the brightness temperature of ALMA 0.87 mm around IRAS4A2.
   The orange line presents the model of one foreground component with $T_{\mbox{\scriptsize dust}} = 20 [K] \times \tau^{0.5}$, where $\tau$ is from 0 to 6.
   The black and gray points in \autoref{fig:PA_sub_BT_Q}(e) and \ref{fig:PA_sub_BT_Q}(f) represent the polarization percentages of the regions corresponded to the white and black line segments shown in \autoref{fig:pol_PI_E}(d), respectively.
   }
   \label{fig:PA_sub_BT_Q}
   }
\end{figure*}

\begin{figure}
    \hspace{-0.2cm}
    \includegraphics[width=9cm]{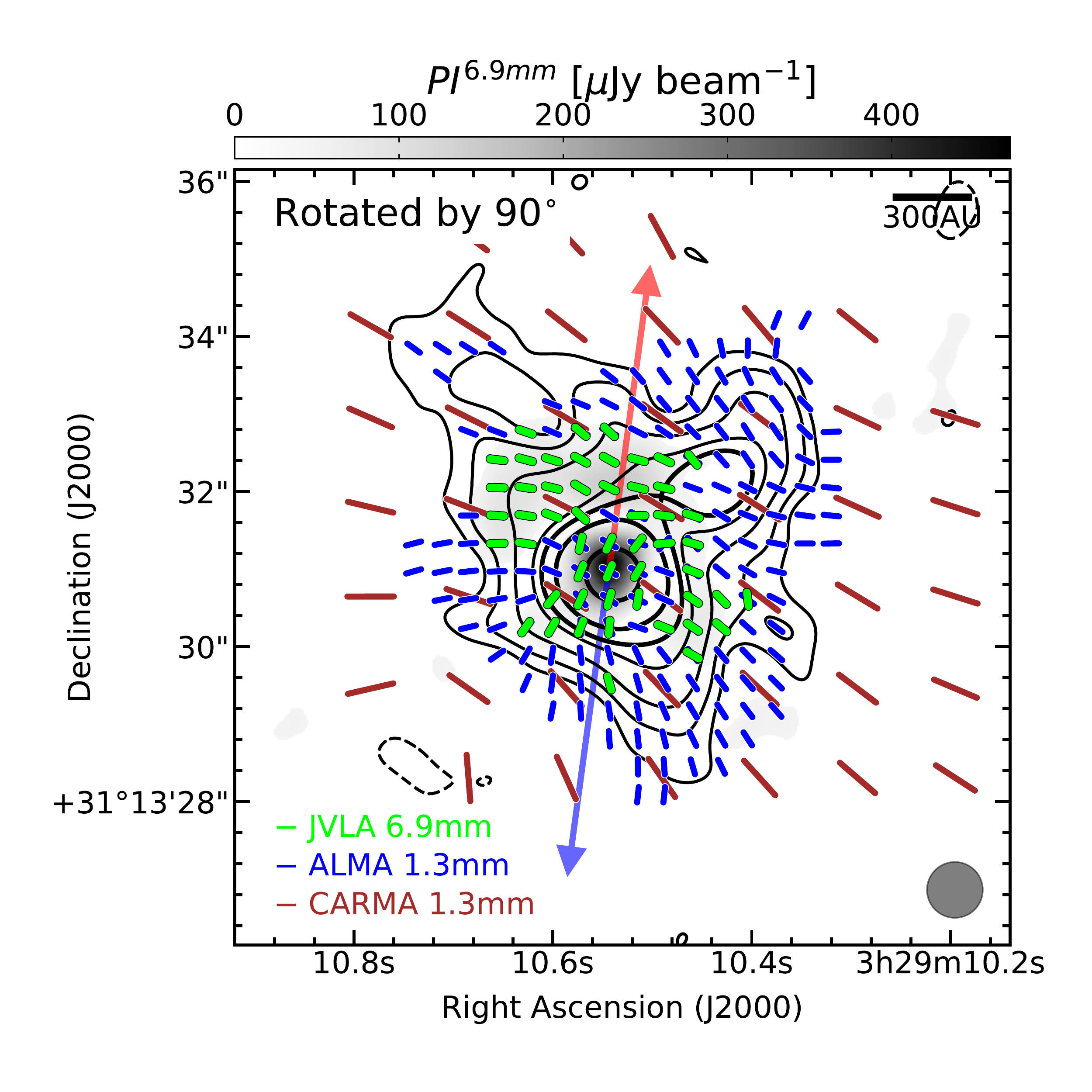}
    \vspace{-0.8cm}
    \caption{\footnotesize{Comparision of position angles (after 90$^\circ$ rotated from the E-field orientation) at 6.9 mm by JVLA (green), at 1.3 mm by ALMA (blue), and 1.3mm by CARMA \citep[brown, ][]{Hull2014}.
    The JVLA and ALMA images presented in this panel were generated using a limited $uv$ distance range of 22-865 $k\lambda$ and were smoothed to the 0\farcs72 resolution.
    Polarization intensity at 6.9 mm is presented in grayscale.
    Contours present the 6.9 mm Stoke I continuum emission.
    Contour levels are 30 $\mu$Jy\,beam (1$\sigma$) $\times$ [-3, 3, 6, 12, 24, 35.5, 71, 284].
    The red and blue arrows represent the axis of $-$9$^\circ$ in the IRAS4A1 redshifted and the blueshifted outflows \citep{Ching2016}, respectively.
    }
    \label{fig:B-field}}
\end{figure}

\acknowledgments 
We thank the anonymous referee for the helpful comments.
This paper makes use of the following ALMA data: ADS/JAO.ALMA \#2015.1.00546.S,  \#2016.1.01089.S. ALMA is a partnership of ESO (representing its member states), NSF (USA) and NINS (Japan), together with NRC (Canada), MOST and ASIAA (Taiwan), and KASI (Republic of Korea), in cooperation with the Republic of Chile. The Joint ALMA Observatory is operated by ESO, AUI/NRAO and NAOJ.
HBL and CLK are supported by the Ministry of Science and Technology (MoST) of Taiwan (Grant No. 108-2112-M-001-002-MY3).
CLK and SPL acknowledge support from the Ministry of Science and Technology of Taiwan with Grant MOST 106-2119-M-007-021-MY3.
JMG is supported by Spanish MINECO AYA2017-84390-C2-1-R grant.

\facility{ALMA, JVLA}
\software{CASA \citep[v5.3.0 + 4.6.0, ][]{McMullin2007}, Numpy \citep{vanderWalt2011}, APLpy \citep{Robitaille2012}
}


\appendix
\section{Polarization measurements}\label{sec:pol_measurements}

\autoref{tab:IRAS4A_value} shows the stokes $I$, $Q$, $U$ intensities, polarization intensity ($PI$), polarization position angle ($PA$), polarization percentage ($P$) and their uncertainties taken at the positions of the polarization segments present in \autoref{fig:pol_PI_E}(a) and \ref{fig:pol_PI_E}(d).  
Details of the observations, data reduction, and imaging are outlined in \autoref{sec:observation}.

\section{Polarization images}\label{sec:pol_images}

\begin{figure*}

   \begin{tabular}{ p{8.5cm} p{8.5cm}}
     \includegraphics[width=8cm]{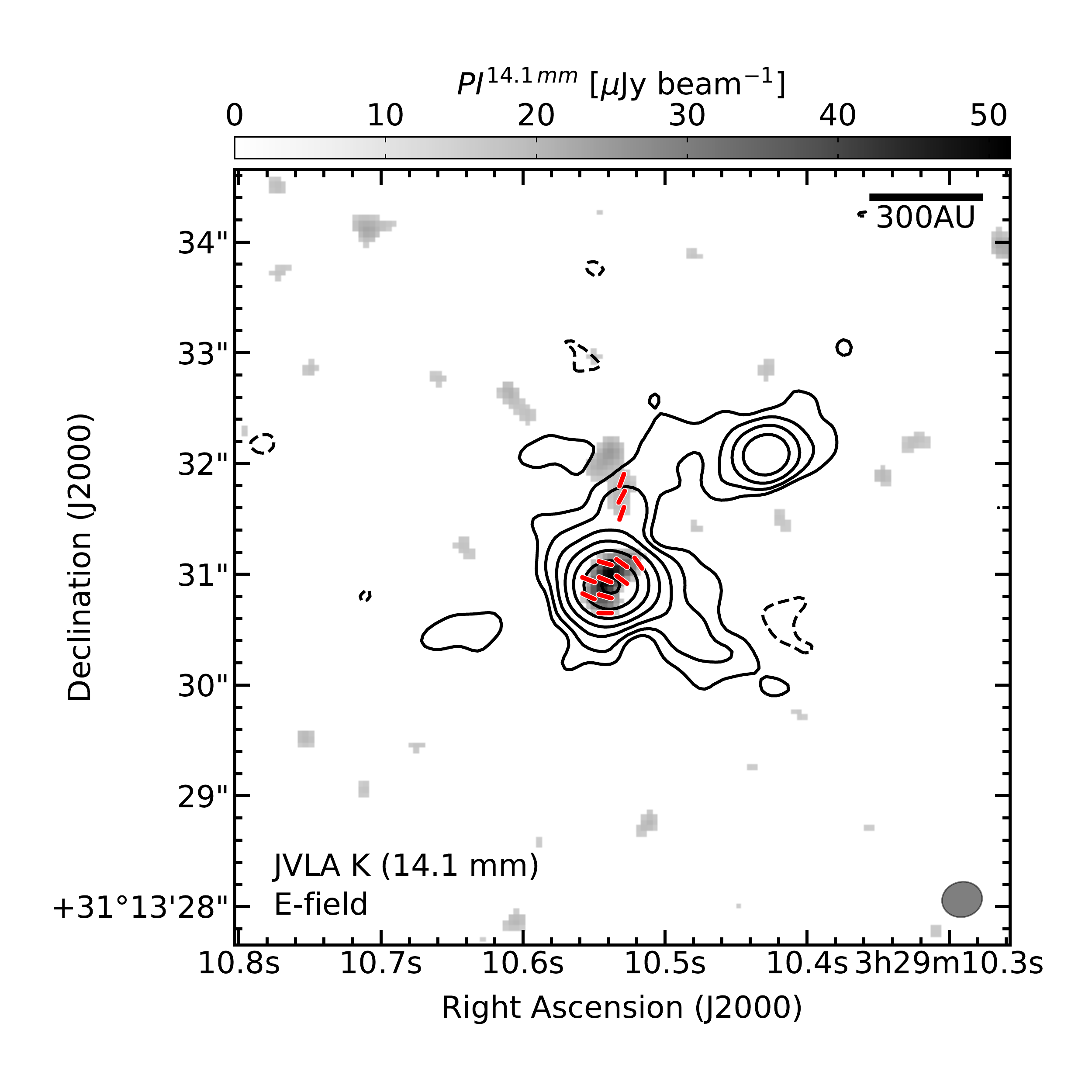} &
	 \includegraphics[width=8cm]{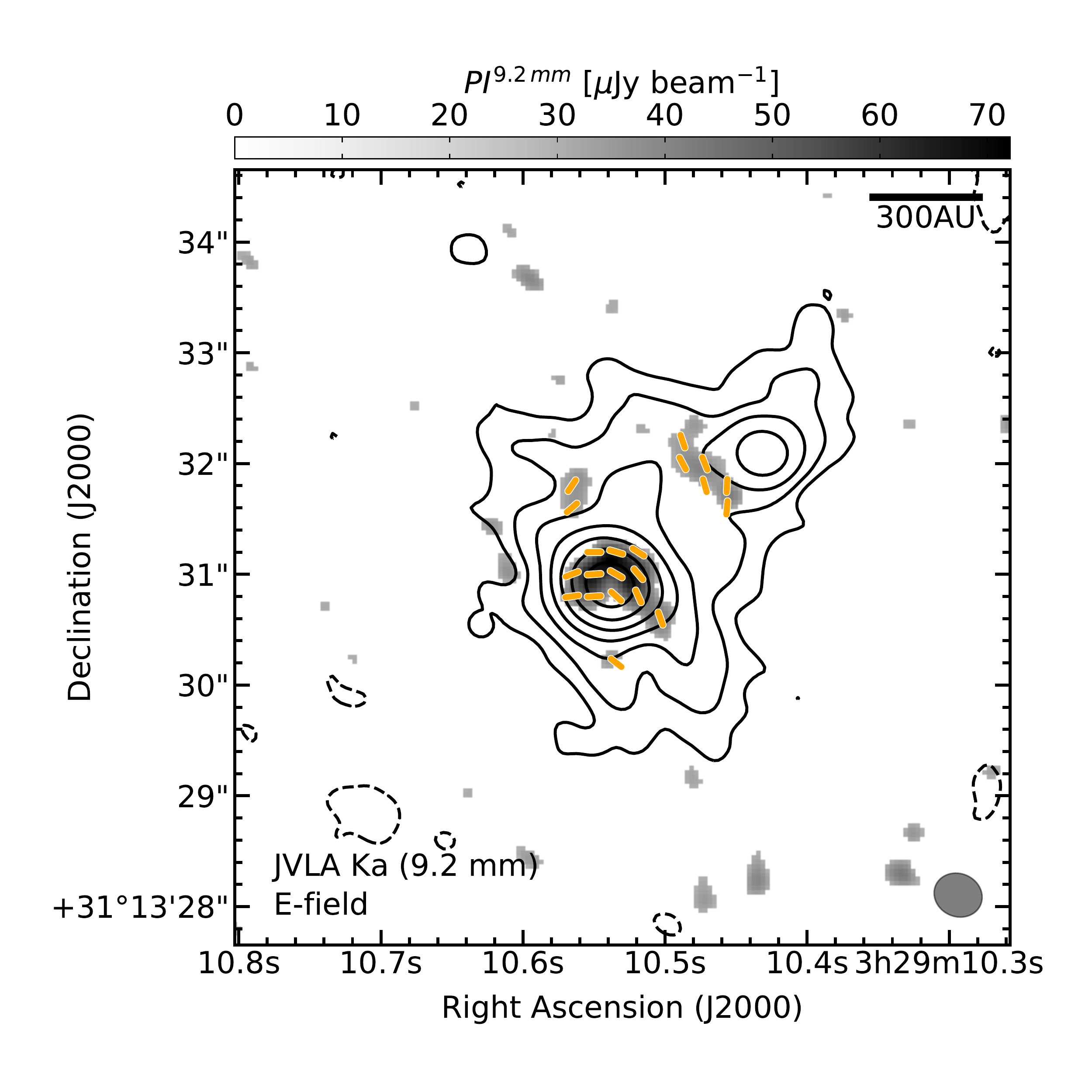}\\
   \end{tabular}
   \vspace{-0.65cm}
   
   \begin{tabular}{ p{8.5cm} p{8.5cm} }
     \includegraphics[width=8cm]{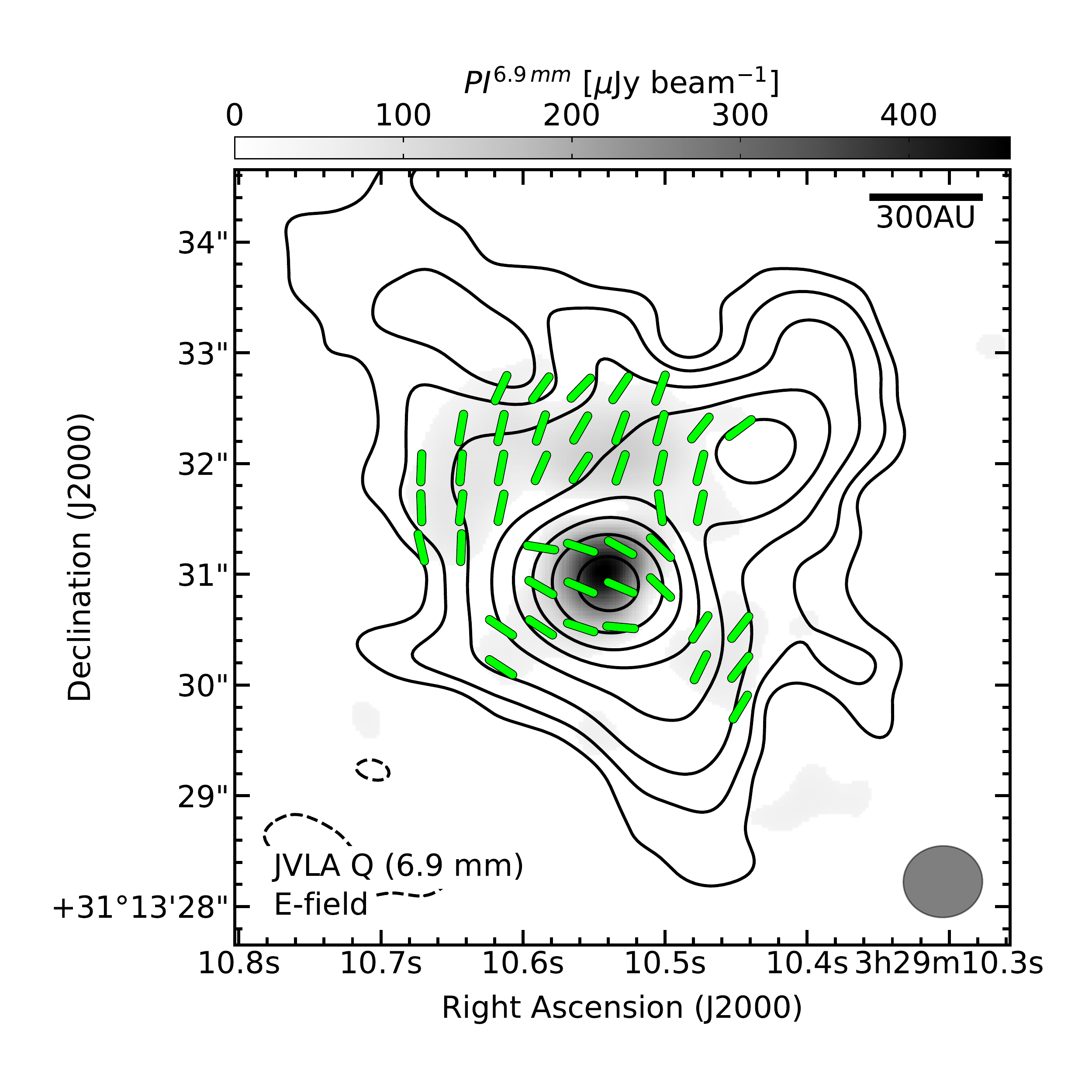} &
	 \includegraphics[width=8cm]{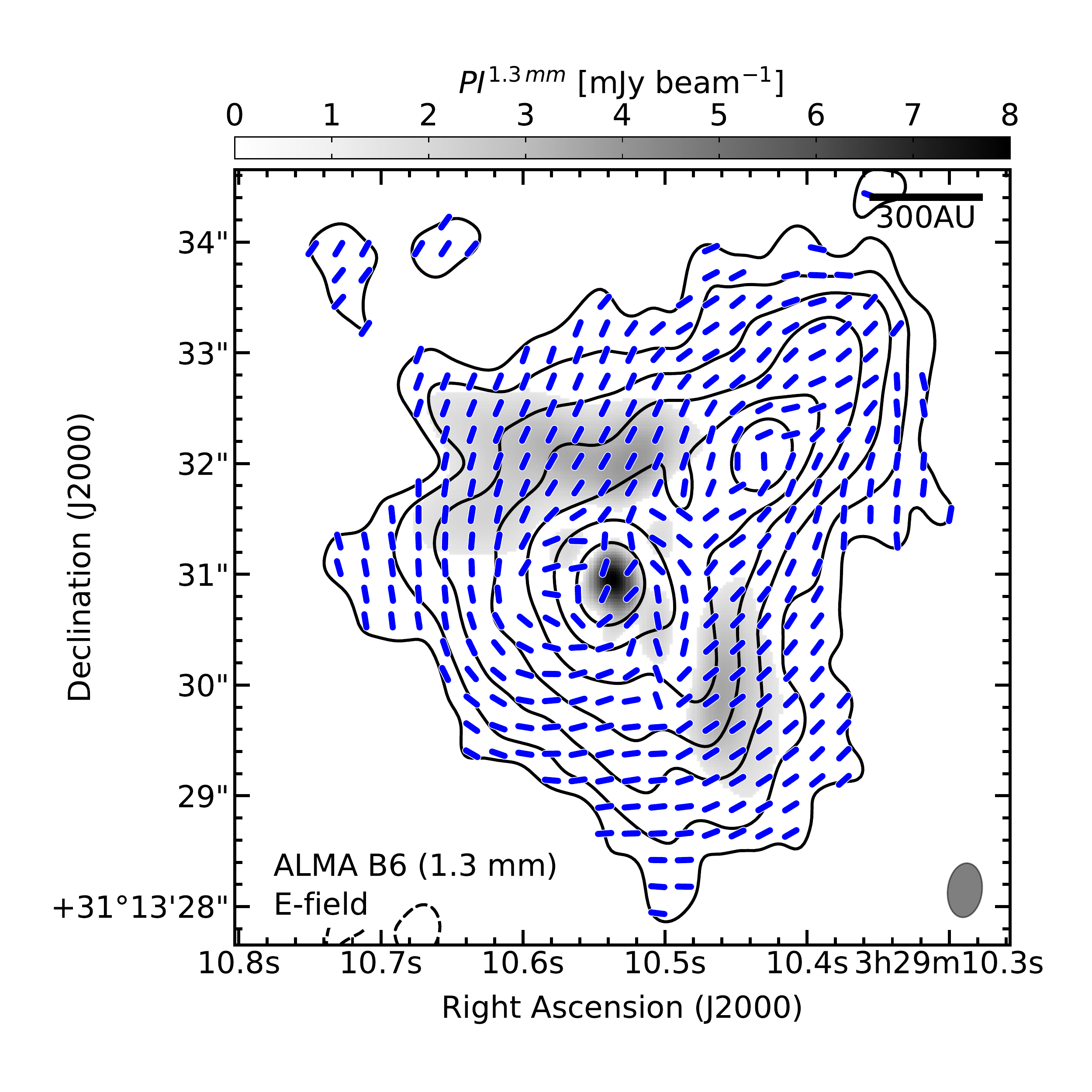} \\
   \end{tabular}
   \vspace{-0.65cm}
   
   \begin{tabular}{ p{8.5cm}}
    \includegraphics[width=8cm]{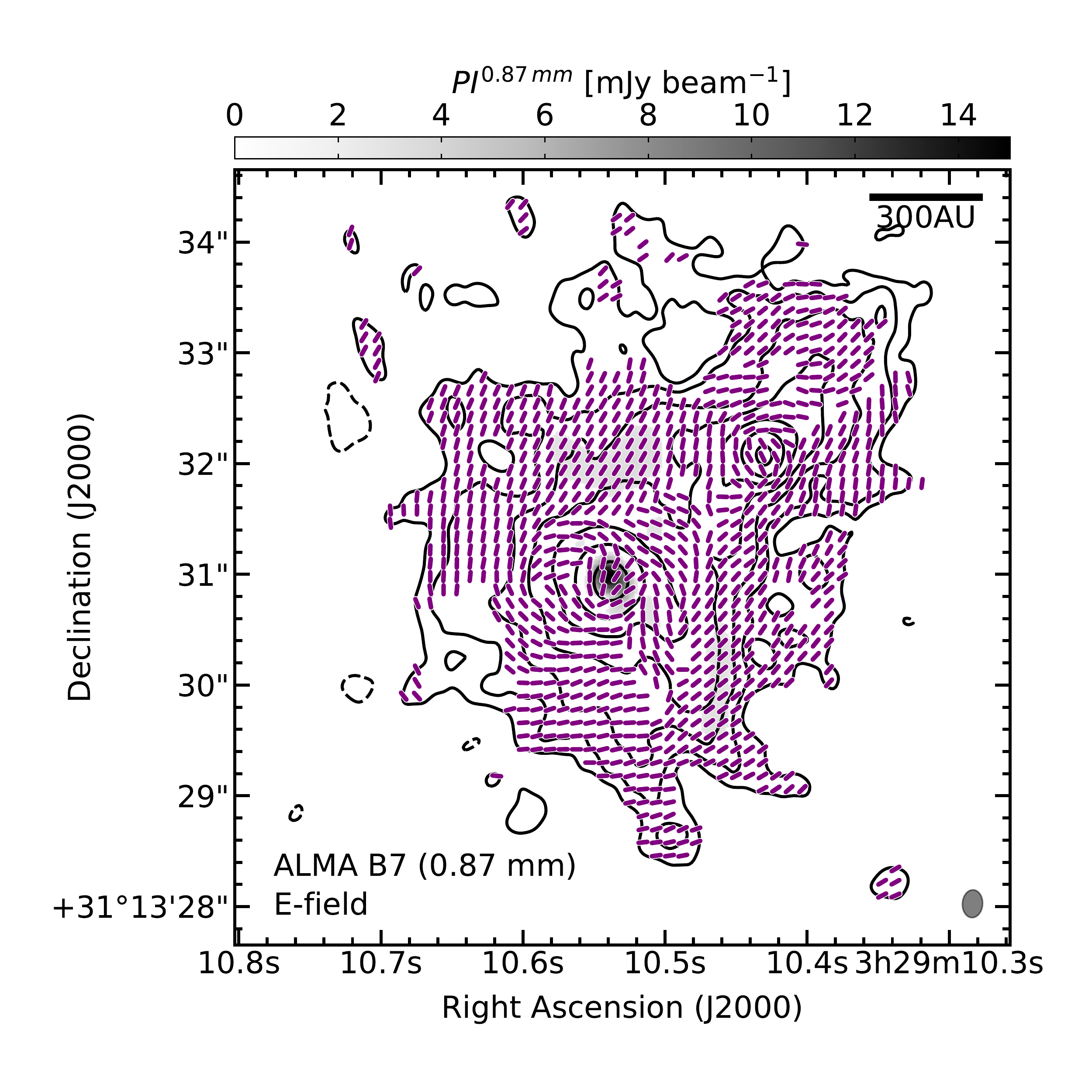}\\
   \end{tabular}
   \vspace{-0.65cm}
   \caption{\footnotesize{
   JVLA K (14.1 mm), Ka (9.2 mm), and Q bands (6.9 mm), and ALMA Band 6 (1.3 mm) and Band 7 (0.87 mm) full polarization observations on NGC1333\,IRAS4A.
   The polarization intensities are presented in grayscale.
   Contours present the stoke I intensities.
   The line segments present the E-field polarization position angles observed at both $>$3 $\sigma_{\mbox{\scriptsize I}}$ and $>$3 $\sigma_{\mbox{\scriptsize PI}}$ significance for each band with the angular resolution listed in \autoref{tab:IRAS4A_obs}.
   The levels of contours are [-1, 1, 2, 4, 8, 16, 32, 64, and 128] $\times$ 3 $\sigma_{\mbox{\scriptsize I}}$, where $\sigma_{\mbox{\scriptsize I}}$ = 5.2, 14, 26, 750, and 1200  $\mu$Jy\,beam$^{-1}$ for the JVLA K, Ka, Q band, and ALMA Band6, and Band 7 stoke I images, respectively.
   All the observations are within a limited $uv$ distance range of 22-865 $k\lambda$.
   }
   \label{fig:pol_PI_E_all}
   }
\end{figure*}

\autoref{fig:pol_PI_E_all} shows the stoke $I$ intensity, polarization intensity ($PI$), and polarization position angle ($PA$) images with JVLA at 11.5--16.7, 8.1--10.3 and 6.3--7.5 mm, and with ALMA at 1.3 and 0.85--0.89 mm towards the Class 0 YSO NGC1333\,IRAS4A.
The images were generated using natural weighting and the limited $uv$ distance range of 22--865 $k\lambda$ with the angular resolution listed in \autoref{tab:IRAS4A_obs}.

\begin{table*}[tpb]
    \caption{Polarization measurements from all the observations \tnote{a}}
    \label{tab:IRAS4A_value}
    \vspace{0.1cm}
    \begin{threeparttable}
    \centering
    \makebox[\linewidth]
    {\footnotesize
    \hspace{-1.4cm}
    \begin{tabular}{ccccccccccc}
        \hline
        \hline \noalign {\vspace{0.05cm}}
    \textbf{\#}\,\tnote{a} & \textbf{RA (J2000)} & \textbf{Dec (J2000)} & \textbf{Stokes I}  & \textbf{Stokes Q} & \textbf{Stokes U}  & \textbf{PI}  & \textbf{PA} & \textbf{$\delta$PA}\,\tnote{b}  & \textbf{P} & \textbf{$\delta$P}\,\tnote{c}  \\
     & (deg) & (deg) & (Jy beam$^{-1}$) & (Jy beam$^{-1}$) & (Jy beam$^{-1}$) & (Jy beam$^{-1}$) & (deg) & (deg) & ($\%$) & ($\%$) \\
    \hline
    \multicolumn{ 11}{c}{\textbf{JVLA K band} \, {(11.5--16.7 mm; 18-26 GHz; beam size: 0\farcs72 $\times$ 0\farcs72; $\sigma_{\mbox{\scriptsize I}}$ $\sim$ 1e-5 Jy beam$^{-1}$; $\sigma_{\mbox{\scriptsize Q}}$ $\sim$ $\sigma_{\mbox{\scriptsize U}}$ $\sim$ $\sigma_{\mbox{\scriptsize PI}}$ $\sim$ 7.4e-6 Jy beam$^{-1}$)}} \\
    \hline 
    1 & 52.29393 & 31.22517 & 1.05e-03 & -2.96e-05 & 2.01e-05 & 3.50e-05 & 73 & 0.1 & 3.3 & 0.7 \\
    1 & 52.29405 & 31.22527 & 6.94e-04 & -1.73e-05 & 2.39e-05 & 2.86e-05 & 63 & 0.13 & 4.1 & 1.1 \\
    1 & 52.29393 & 31.22527 & 1.57e-03 & -3.66e-05 & 4.06e-05 & 5.42e-05 & 66 & 0.068 & 3.4 & 0.5 \\
    1 & 52.29381 & 31.22527 & 1.14e-03 & -1.86e-05 & 2.31e-05 & 2.87e-05 & 64 & 0.12 & 2.5 & 0.65 \\
    1 & 52.29393 & 31.22547 & 1.87e-04 & 2.95e-05 & -1.27e-05 & 3.12e-05 & -12 & 0.12 & 17 & 4.1 \\
    1 & 52.29393 & 31.22557 & 6.51e-05 & 3.50e-05 & -1.52e-05 & 3.75e-05 & -12 & 0.097 & 58 & 15 \\
    \hline
    \multicolumn{ 11}{c}{\textbf{JVLA Ka band} \, {(8.1--10 mm; 29-37 GHz; beam size: 0\farcs72 $\times$ 0\farcs72; $\sigma_{\mbox{\scriptsize I}}$ $\sim$ 4e-5 Jy beam$^{-1}$; $\sigma_{\mbox{\scriptsize Q}}$ $\sim$ $\sigma_{\mbox{\scriptsize U}}$ $\sim$ $\sigma_{\mbox{\scriptsize PI}}$ $\sim$ 1.7e-5 Jy beam$^{-1}$)}} \\
    \hline
    2 & 52.29381 & 31.22517 & 2.81e-03 & 2.56e-05 & 6.32e-05 & 6.60e-05 & 34 & 0.12 & 2.4 & 0.61 \\
    2 & 52.29393 & 31.22527 & 5.55e-03 & -7.31e-05 & 3.20e-05 & 7.79e-05 & 78 & 0.11 & 1.4 & 0.5 \\
    2 & 52.29381 & 31.22527 & 4.01e-03 & -1.54e-06 & 7.52e-05 & 7.33e-05 & 46 & 0.11 & 1.8 & 0.5 \\
    2 & 52.29405 & 31.22547 & 4.93e-04 & 1.20e-05 & -5.56e-05 & 5.42e-05 & -39 & 0.15 & 11 & 3.6 \\
    2 & 52.29358 & 31.22547 & 4.23e-04 & 5.39e-05 & -8.79e-07 & 5.12e-05 & -0 & 0.16 & 12 & 4.2 \\
    \multicolumn{ 11}{c}{...}\\
    \hline
    \multicolumn{ 11}{c}{\textbf{JVLA Q band} \, {(6.3--7.5 mm; 40-48 GHz; beam size: 0\farcs72 $\times$ 0\farcs72; $\sigma_{\mbox{\scriptsize I}}$ $\sim$ 3e-5 Jy beam$^{-1}$; $\sigma_{\mbox{\scriptsize Q}}$ $\sim$ $\sigma_{\mbox{\scriptsize U}}$ $\sim$ $\sigma_{\mbox{\scriptsize PI}}$ $\sim$ 1.4e-5 Jy beam$^{-1}$)}} \\
    \hline
    3 & 52.29393 & 31.22487 & 2.69e-04 & -4.73e-05 & -2.97e-05 & 5.41e-05 & -74 & 0.13 & 20 & 5.7 \\
    3 & 52.29358 & 31.22497 & 5.27e-04 & 3.07e-05 & -5.19e-05 & 5.87e-05 & -30 & 0.12 & 11 & 2.7 \\
    3 & 52.29428 & 31.22507 & 3.63e-04 & -1.71e-05 & 4.52e-05 & 4.62e-05 & 55 & 0.14 & 13 & 4 \\
    3 & 52.29417 & 31.22507 & 6.17e-04 & -2.96e-05 & 5.16e-05 & 5.78e-05 & 60 & 0.12 & 9.4 & 2.3 \\
    3 & 52.29405 & 31.22507 & 1.30e-03 & -4.22e-05 & 3.96e-05 & 5.61e-05 & 68 & 0.12 & 4.3 & 1.1 \\
    \multicolumn{ 11}{c}{...}\\
    \hline
    \multicolumn{ 11}{c}{\textbf{ALMA Band 6} \, {(1.3 mm; 234 GHz; beam size: 0\farcs72 $\times$ 0\farcs72; $\sigma_{\mbox{\scriptsize I}}$ $\sim$ 2e-3 Jy beam$^{-1}$; $\sigma_{\mbox{\scriptsize Q}}$ $\sim$ $\sigma_{\mbox{\scriptsize U}}$ $\sim$ $\sigma_{\mbox{\scriptsize PI}}$ $\sim$ 1.2e-4 Jy beam$^{-1}$)}} \\
    \hline
    4 & 52.29381 & 31.22447 & 6.75e-03 & -1.15e-03 & 2.36e-04 & 1.17e-03 & 84 & 0.051 & 17 & 5.5 \\
    4 & 52.29370 & 31.22447 & 7.27e-03 & -1.09e-03 & 2.13e-04 & 1.10e-03 & 84 & 0.054 & 15 & 4.5 \\
    4 & 52.29381 & 31.22457 & 1.05e-02 & -1.50e-03 & -4.47e-05 & 1.50e-03 & -89 & 0.04 & 14 & 3 \\
    4 & 52.29370 & 31.22457 & 1.14e-02 & -1.29e-03 & -1.64e-04 & 1.30e-03 & -86 & 0.046 & 11 & 2.3 \\
    4 & 52.29358 & 31.22457 & 8.89e-03 & -1.01e-03 & -6.21e-04 & 1.18e-03 & -74 & 0.051 & 13 & 3.3 \\
    \multicolumn{ 11}{c}{...}\\
    \hline
    \multicolumn{ 11}{c}{\textbf{ALMA Band 7} \, {(0.87 mm; 345 GHz; beam size: 0\farcs72 $\times$ 0\farcs72; $\sigma_{\mbox{\scriptsize I}}$ $\sim$ 9e-3 Jy beam$^{-1}$; $\sigma_{\mbox{\scriptsize Q}}$ $\sim$ $\sigma_{\mbox{\scriptsize U}}$ $\sim$ $\sigma_{\mbox{\scriptsize PI}}$ $\sim$ 8e-4 Jy beam$^{-1}$)}} \\
    \hline
    5 & 52.29370 & 31.22457 & 3.20e-02 & -3.70e-03 & -2.54e-03 & 4.42e-03 & -73 & 0.089 & 14 & 4.7 \\
    5 & 52.29381 & 31.22467 & 4.80e-02 & -4.53e-03 & -2.18e-03 & 4.96e-03 & -77 & 0.08 & 10 & 2.6 \\
    5 & 52.29370 & 31.22467 & 4.76e-02 & -4.35e-03 & -3.31e-03 & 5.41e-03 & -71 & 0.073 & 11 & 2.7 \\
    5 & 52.29358 & 31.22467 & 2.94e-02 & -4.90e-03 & -5.52e-03 & 7.34e-03 & -66 & 0.054 & 25 & 8.2 \\
    5 & 52.29417 & 31.22477 & 2.88e-02 & -3.48e-03 & 6.53e-05 & 3.39e-03 & 89 & 0.11 & 12 & 4.7 \\
    \multicolumn{ 11}{c}{...}\\
    \hline
    \multicolumn{ 11}{c}{\textbf{ALMA Band 7} \, {(0.87 mm; 345 GHz; beam size: 0\farcs25 $\times$ 0\farcs18; $\sigma_{\mbox{\scriptsize I}}$ $\sim$ 1.2e-3 Jy beam$^{-1}$; $\sigma_{\mbox{\scriptsize Q}}$ $\sim$ $\sigma_{\mbox{\scriptsize U}}$ $\sim$ $\sigma_{\mbox{\scriptsize PI}}$ $\sim$ 9e-5 Jy beam$^{-1}$)}} \\
    \hline
    6 & 52.293113 & 31.224472 & 0.0037 & -0.0002 & -0.00027 & 0.00033 & -63 & 0.13 & 8.8 & 3.8 \\
    6 & 52.293074 & 31.224472 & 0.0041 & -0.0002 & -0.00022 & 0.00029 & -66 & 0.15 & 6.9 & 3 \\
    6 & 52.293113 & 31.224505 & 0.005 & -0.00017 & -0.00032 & 0.00035 & -59 & 0.12 & 7.1 & 2.5 \\
    6 & 52.293074 & 31.224505 & 0.0055 & -0.0002 & -0.00031 & 0.00035 & -61 & 0.12 & 6.4 & 2.2 \\
    6 & 52.293074 & 31.224539 & 0.0039 & -0.0002 & -0.00034 & 0.00039 & -60 & 0.11 & 10 & 3.9 \\
    \multicolumn{ 11}{c}{...}\\
    \hline \noalign {\vspace{0.05cm}}
    \end{tabular}
    }
    \begin{tablenotes}
        \item \textbf{Notes.}
        All the measurements are from the images of limited $uv$ distance range of 22--865 k$\lambda$ and naturally weighted.
        \item[a] Each number represents the following observation:
        \item (1): JVLA K band (beam size: 0\farcs72$\times$0\farcs72)
        (2): JVLA Ka band (beam size: 0\farcs72$\times$0\farcs72)
        (3): JVLA Q band (beam size: 0\farcs72$\times$0\farcs72)
        \item  (4): ALMA Band 6 (beam size: 0\farcs72$\times$0\farcs72)
        (5): ALMA Band 7 (beam size: 0\farcs72$\times$0\farcs72)
        (6): ALMA Band 7 (beam size: 0\farcs25$\times$0\farcs18)
        \item[b] The uncertainty of polarization position angle does not take into account the systematic calibration uncertainty.
        \item[c] We have considered the systematic calibration uncertainties in polarization percentages, which is 0.5\% and 0.1\% for the JVLA and the ALMA observations, respectively, according to the Guide to Observing with the VLA (\href{https://science.nrao.edu/facilities/vla/docs/manuals/obsguide/modes/pol}{https://science.nrao.edu/facilities/vla/docs/manuals/obsguide/modes/pol}) and the ALMA Cycle 3 and Cycle 4 Technical Handbook (\href{https://almascience.nrao.edu/documents-and-tools/cycle4/alma-technical-handbook/view}{https://almascience.nrao.edu/documents-and-tools/cycle4/alma-technical-handbook/view}).
        \item \autoref{tab:IRAS4A_value} is published in its entirety in the machine-readable format.
        A portion is shown here for guidance regarding its form and content.

    \end{tablenotes}
    \end{threeparttable}
\end{table*}


\begin{thebibliography}{}
\providecommand\natexlab[1]{#1}
\providecommand\JournalTitle[1]{#1}

\bibitem[{{Akeson} \& {Carlstrom}(1997)}]{Akeson1997}
{Akeson}, R.~L., \& {Carlstrom}, J.~E. 1997,
  \href{http://dx.doi.org/10.1086/304949}{\JournalTitle{\apj}, 491, 254}

\bibitem[{{Akeson} {et~al.}(1996){Akeson}, {Carlstrom}, {Phillips}, \&
  {Woody}}]{Akeson1996}
{Akeson}, R.~L., {Carlstrom}, J.~E., {Phillips}, J.~A., \& {Woody}, D.~P. 1996,
  \href{http://dx.doi.org/10.1086/309856}{\JournalTitle{\apjl}, 456, L45}

\bibitem[{{Ching} {et~al.}(2016){Ching}, {Lai}, {Zhang}, {Yang}, {Girart}, \&
  {Rao}}]{Ching2016}
{Ching}, T.-C., {Lai}, S.-P., {Zhang}, Q., {et~al.} 2016,
  \href{http://dx.doi.org/10.3847/0004-637X/819/2/159}{\JournalTitle{\apj},
  819, 159}

\bibitem[{{Cox} {et~al.}(2015){Cox}, {Harris}, {Looney}, {Segura-Cox}, {Tobin},
  {Li}, {Tychoniec}, {Chandler}, {Dunham}, {Kratter}, {Melis}, {Perez}, \&
  {Sadavoy}}]{Cox2015}
{Cox}, E.~G., {Harris}, R.~J., {Looney}, L.~W., {et~al.} 2015,
  \href{http://dx.doi.org/10.1088/2041-8205/814/2/L28}{\JournalTitle{\apjl},
  814, L28}

\bibitem[{{Crutcher}(2012)}]{Crutcher2012}
{Crutcher}, R.~M. 2012,
  \href{http://dx.doi.org/10.1146/annurev-astro-081811-125514}{\JournalTitle{\araa},
  50, 29}

\bibitem[{{Dent} {et~al.}(2019){Dent}, {Pinte}, {Cortes}, {M{\'e}nard},
  {Hales}, {Fomalont}, \& {de Gregorio-Monsalvo}}]{Dent2019}
{Dent}, W.~R.~F., {Pinte}, C., {Cortes}, P.~C., {et~al.} 2019,
  \href{http://dx.doi.org/10.1093/mnrasl/sly181}{\JournalTitle{\mnras}, 482,
  L29}

\bibitem[{{Frank} {et~al.}(2014){Frank}, {Ray}, {Cabrit}, {Hartigan}, {Arce},
  {Bacciotti}, {Bally}, {Benisty}, {Eisl{\"o}ffel}, {G{\"u}del}, {Lebedev},
  {Nisini}, \& {Raga}}]{Frank2014}
{Frank}, A., {Ray}, T.~P., {Cabrit}, S., {et~al.} 2014,
  \href{http://dx.doi.org/10.2458/azu_uapress_9780816531240-ch020}{\JournalTitle{Protostars
  and Planets VI}, 451}

\bibitem[{{Frau} {et~al.}(2011){Frau}, {Galli}, \& {Girart}}]{Frau2011}
{Frau}, P., {Galli}, D., \& {Girart}, J.~M. 2011,
  \href{http://dx.doi.org/10.1051/0004-6361/201117813}{\JournalTitle{\aap},
  535, A44}

\bibitem[{{Galametz} {et~al.}(2018){Galametz}, {Maury}, {Girart}, {Rao},
  {Zhang}, {Gaudel}, {Valdivia}, {Keto}, \& {Lai}}]{Galametz2018}
{Galametz}, M., {Maury}, A., {Girart}, J.~M., {et~al.} 2018,
  \href{http://dx.doi.org/10.1051/0004-6361/201833004}{\JournalTitle{\aap},
  616, A139}

\bibitem[{{Galv{\'a}n-Madrid} {et~al.}(2018){Galv{\'a}n-Madrid}, {Liu},
  {Izquierdo}, {Miotello}, {Zhao}, {Carrasco-Gonz{\'a}lez}, {Lizano}, \&
  {Rodr{\'\i}guez}}]{Galvan-Madrid2018}
{Galv{\'a}n-Madrid}, R., {Liu}, H.~B., {Izquierdo}, A.~F., {et~al.} 2018,
  \href{http://dx.doi.org/10.3847/1538-4357/aae779}{\JournalTitle{\apj}, 868,
  39}

\bibitem[{{Girart} {et~al.}(1999){Girart}, {Crutcher}, \& {Rao}}]{Girart1999}
{Girart}, J.~M., {Crutcher}, R.~M., \& {Rao}, R. 1999,
  \href{http://dx.doi.org/10.1086/312345}{\JournalTitle{\apjl}, 525, L109}

\bibitem[{{Girart} {et~al.}(2006){Girart}, {Rao}, \& {Marrone}}]{Girart2006}
{Girart}, J.~M., {Rao}, R., \& {Marrone}, D.~P. 2006,
  \href{http://dx.doi.org/10.1126/science.1129093}{\JournalTitle{Science}, 313,
  812}

\bibitem[{{Gon{\c c}alves} {et~al.}(2008){Gon{\c c}alves}, {Galli}, \&
  {Girart}}]{Goncalves2008}
{Gon{\c c}alves}, J., {Galli}, D., \& {Girart}, J.~M. 2008,
  \href{http://dx.doi.org/10.1051/0004-6361:200810861}{\JournalTitle{\aap},
  490, L39}

\bibitem[{{Hildebrand} {et~al.}(2000){Hildebrand}, {Davidson}, {Dotson},
  {Dowell}, {Novak}, \& {Vaillancourt}}]{Hildebrand2000}
{Hildebrand}, R.~H., {Davidson}, J.~A., {Dotson}, J.~L., {et~al.} 2000,
  \href{http://dx.doi.org/10.1086/316613}{\JournalTitle{\pasp}, 112, 1215}

\bibitem[{{Hull} {et~al.}(2014){Hull}, {Plambeck}, {Kwon}, {Bower},
  {Carpenter}, {Crutcher}, {Fiege}, {Franzmann}, {Hakobian}, {Heiles}, {Houde},
  {Hughes}, {Lamb}, {Looney}, {Marrone}, {Matthews}, {Pillai}, {Pound},
  {Rahman}, {Sandell}, {Stephens}, {Tobin}, {Vaillancourt}, {Volgenau}, \&
  {Wright}}]{Hull2014}
{Hull}, C.~L.~H., {Plambeck}, R.~L., {Kwon}, W., {et~al.} 2014,
  \href{http://dx.doi.org/10.1088/0067-0049/213/1/13}{\JournalTitle{\apjs},
  213, 13}

\bibitem[{{Hull} {et~al.}(2018){Hull}, {Yang}, {Li}, {Kataoka}, {Stephens},
  {Andrews}, {Bai}, {Cleeves}, {Hughes}, {Looney}, {P{\'e}rez}, \&
  {Wilner}}]{Hull2018}
{Hull}, C.~L.~H., {Yang}, H., {Li}, Z.-Y., {et~al.} 2018,
  \href{http://dx.doi.org/10.3847/1538-4357/aabfeb}{\JournalTitle{\apj}, 860,
  82}

\bibitem[{{Kataoka} {et~al.}(2017){Kataoka}, {Tsukagoshi}, {Pohl}, {Muto},
  {Nagai}, {Stephens}, {Tomisaka}, \& {Momose}}]{Kataoka2017}
{Kataoka}, A., {Tsukagoshi}, T., {Pohl}, A., {et~al.} 2017,
  \href{http://dx.doi.org/10.3847/2041-8213/aa7e33}{\JournalTitle{\apj}, 844,
  L5}

\bibitem[{{Kataoka} {et~al.}(2015){Kataoka}, {Muto}, {Momose}, {Tsukagoshi},
  {Fukagawa}, {Shibai}, {Hanawa}, {Murakawa}, \& {Dullemond}}]{Kataoka2015}
{Kataoka}, A., {Muto}, T., {Momose}, M., {et~al.} 2015,
  \href{http://dx.doi.org/10.1088/0004-637X/809/1/78}{\JournalTitle{\apj}, 809,
  78}
  
\bibitem[{{Kataoka} {et~al.}(2016){Kataoka}, {Muto}, {Momose}, {Tsukagoshi}, \&
  {Dullemond}}]{Kataoka2016}
{Kataoka}, A., {Muto}, T., {Momose}, M., {Tsukagoshi}, T., \& {Dullemond},
  C.~P. 2016,
  \href{http://dx.doi.org/10.3847/0004-637X/820/1/54}{\JournalTitle{\apj}, 820,
  54}

\bibitem[{{Li} {et~al.}(2017){Li}, {Liu}, {Hasegawa}, \& {Hirano}}]{Li2017}
{Li}, J. I.-H., {Liu}, H.~B., {Hasegawa}, Y., \& {Hirano}, N. 2017,
  \href{http://dx.doi.org/10.3847/1538-4357/aa6f04}{\JournalTitle{\apj}, 840,
  72}

\bibitem[{{Li} {et~al.}(2014){Li}, {Krasnopolsky}, {Shang}, \& {Zhao}}]{Li2014}
{Li}, Z.-Y., {Krasnopolsky}, R., {Shang}, H., \& {Zhao}, B. 2014,
  \href{http://dx.doi.org/10.1088/0004-637X/793/2/130}{\JournalTitle{\apj},
  793, 130}

\bibitem[{{Liu} {et~al.}(2018){Liu}, {Hasegawa}, {Ching}, {Lai}, {Hirano}, \&
  {Rao}}]{Liu2018}
{Liu}, H.~B., {Hasegawa}, Y., {Ching}, T.-C., {et~al.} 2018,
  \href{http://dx.doi.org/10.1051/0004-6361/201832699}{\JournalTitle{\aap},
  617, A3}

\bibitem[{{Liu} {et~al.}(2016){Liu}, {Lai}, {Hasegawa}, {Hirano}, {Rao}, {Li},
  {Fukagawa}, {Girart}, {Carrasco-Gonz{\'a}lez}, \&
  {Rodr{\'{\i}}guez}}]{Liu2016}
{Liu}, H.~B., {Lai}, S.-P., {Hasegawa}, Y., {et~al.} 2016,
  \href{http://dx.doi.org/10.3847/0004-637X/821/1/41}{\JournalTitle{\apj}, 821,
  41}
  
\bibitem[{{Liu}(2019)}]{Liu2019ApJ}
{Liu}, H.~B. 2019,
  \href{http://dx.doi.org/10.3847/2041-8213/ab1f8e}{\JournalTitle{The
  Astrophysical Journal}, 877, L22}

\bibitem[{{McMullin} {et~al.}(2007){McMullin}, {Waters}, {Schiebel}, {Young},
  \& {Golap}}]{McMullin2007}
{McMullin}, J.~P., {Waters}, B., {Schiebel}, D., {Young}, W., \& {Golap}, K.
  2007, in Astronomical Society of the Pacific Conference Series, Vol. 376,
  Astronomical Data Analysis Software and Systems XVI, ed. R.~A. {Shaw},
  F.~{Hill}, \& D.~J. {Bell}, 127

\bibitem[{{Mouschovias}(1977)}]{Mouschovias1977}
{Mouschovias}, T.~C. 1977,
  \href{http://dx.doi.org/10.1086/154912}{\JournalTitle{\apj}, 211, 147}

\bibitem[{{Ohashi} {et~al.}(2018){Ohashi}, {Kataoka}, {Nagai}, {Momose},
  {Muto}, {Hanawa}, {Fukagawa}, {Tsukagoshi}, {Murakawa}, \&
  {Shibai}}]{Ohashi2018}
{Ohashi}, S., {Kataoka}, A., {Nagai}, H., {et~al.} 2018,
  \href{http://dx.doi.org/10.3847/1538-4357/aad632}{\JournalTitle{\apj}, 864,
  81}

\bibitem[{{Okuzumi} \& {Tazaki}(2019)}]{Okuzumi2019}
{Okuzumi}, S., \& {Tazaki}, R. 2019,
  \href{http://dx.doi.org/10.3847/1538-4357/ab204d}{\JournalTitle{\apj}, 878,
  132}

\bibitem[{{Ortiz-Le{\'o}n} {et~al.}(2018){Ortiz-Le{\'o}n}, {Loinard}, {Dzib},
  {Galli}, {Kounkel}, {Mioduszewski}, {Rodr{\'{\i}}guez}, {Torres}, {Hartmann},
  {Boden}, {Evans}, {Brice{\~n}o}, \& {Tobin}}]{Ortiz2018}
{Ortiz-Le{\'o}n}, G.~N., {Loinard}, L., {Dzib}, S.~A., {et~al.} 2018,
  \href{http://dx.doi.org/10.3847/1538-4357/aada49}{\JournalTitle{\apj}, 865,
  73}
  
\bibitem[{{Robitaille} \& {Bressert}(2012)}]{Robitaille2012}
{Robitaille}, T., \& {Bressert}, E. 2012, {APLpy: Astronomical Plotting Library
  in Python}

\bibitem[{{Sahu} {et~al.}(2019){Sahu}, {Liu}, {Su}, {Li}, {Lee}, {Hirano}, \&
  {Takakuwa}}]{Sahu2019}
{Sahu}, D., {Liu}, S.-Y., {Su}, Y.-N., {et~al.} 2019,
  \href{http://dx.doi.org/10.3847/1538-4357/aaffda}{\JournalTitle{\apj}, 872,
  196}

\bibitem[{{Santangelo} {et~al.}(2015){Santangelo}, {Codella}, {Cabrit},
  {Maury}, {Gueth}, {Maret}, {Lefloch}, {Belloche}, {Andr{\'e}}, {Hennebelle},
  {Anderl}, {Podio}, \& {Testi}}]{Santangelo2015}
{Santangelo}, G., {Codella}, C., {Cabrit}, S., {et~al.} 2015,
  \href{http://dx.doi.org/10.1051/0004-6361/201526323}{\JournalTitle{\aap},
  584, A126}

\bibitem[{{Shu}(1977)}]{Shu1977ApJ}
{Shu}, F.~H. 1977, \href{http://dx.doi.org/10.1086/155274}{\JournalTitle{\apj},
  214, 488}

\bibitem[{{Shu} {et~al.}(1987){Shu}, {Adams}, \& {Lizano}}]{Shu1987}
{Shu}, F.~H., {Adams}, F.~C., \& {Lizano}, S. 1987,
  \href{http://dx.doi.org/10.1146/annurev.aa.25.090187.000323}{\JournalTitle{\araa},
  25, 23}

\bibitem[{{Simmons} \& {Stewart}(1985)}]{Simmons1985}
{Simmons}, J.~F.~L., \& {Stewart}, B.~G. 1985, \JournalTitle{\aap}, 142, 100

\bibitem[{{Stephens} {et~al.}(2017){Stephens}, {Yang}, {Li}, {Looney},
  {Kataoka}, {Kwon}, {Fern{\'a}ndez-L{\'o}pez}, {Hull}, {Hughes}, {Segura-Cox},
  {Mundy}, {Crutcher}, \& {Rao}}]{Stephens2017}
{Stephens}, I.~W., {Yang}, H., {Li}, Z.-Y., {et~al.} 2017,
  \href{http://dx.doi.org/10.3847/1538-4357/aa998b}{\JournalTitle{\apj}, 851,
  55}
  
\bibitem[{{Tazaki} {et~al.}(2019){Tazaki}, {Tanaka}, {Kataoka}, {Okuzumi}, \&
  {Muto}}]{Takaki2019}
{Tazaki}, R., {Tanaka}, H., {Kataoka}, A., {Okuzumi}, S., \& {Muto}, T. 2019,
  arXiv e-prints, arXiv:1907.00189  
  
\bibitem[{{Testi} {et~al.}(2014){Testi}, {Birnstiel}, {Ricci}, {Andrews},
  {Blum}, {Carpenter}, {Dominik}, {Isella}, {Natta}, {Williams}, \&
  {Wilner}}]{Testi2014}
{Testi}, L., {Birnstiel}, T., {Ricci}, L., {et~al.} 2014,
  \href{http://dx.doi.org/10.2458/azu_uapress_9780816531240-ch015}{in
  Protostars and Planets VI, ed. H.~{Beuther}, R.~S. {Klessen}, C.~P.
  {Dullemond}, \& T.~{Henning}}, 339

\bibitem[{{Vaillancourt}(2006)}]{Vaillancourt2006}
{Vaillancourt}, J.~E. 2006,
  \href{http://dx.doi.org/10.1086/507472}{\JournalTitle{\pasp}, 118, 1340}
  
\bibitem[{{van der Walt} {et~al.}(2011){van der Walt}, {Colbert}, \&
  {Varoquaux}}]{vanderWalt2011}
{van der Walt}, S., {Colbert}, S.~C., \& {Varoquaux}, G. 2011,
  \href{http://dx.doi.org/10.1109/MCSE.2011.37}{\JournalTitle{Computing in
  Science and Engineering}, 13, 22}

\bibitem[{{Yang} {et~al.}(2016){Yang}, {Li}, {Looney}, {Cox}, {Tobin},
  {Stephens}, {Segura-Cox}, \& {Harris}}]{Yang2016}
{Yang}, H., {Li}, Z.-Y., {Looney}, L.~W., {et~al.} 2016,
  \href{http://dx.doi.org/10.1093/mnras/stw1253}{\JournalTitle{\mnras}, 460,
  4109}

\bibitem[{{Zhu} {et~al.}(2019){Zhu}, {Zhang}, {Jiang}, {Kataoka}, {Birnstiel},
  {Dullemond}, {Andrews}, {Huang}, {P{\'e}rez}, {Carpenter}, {Bai}, {Wilner},
  \& {Ricci}}]{Zhu2019ApJ}
{Zhu}, Z., {Zhang}, S., {Jiang}, Y.-F., {et~al.} 2019,
  \href{http://dx.doi.org/10.3847/2041-8213/ab1f8c}{\JournalTitle{The
  Astrophysical Journal}, 877, L18}

\bibitem[{{Zucker} {et~al.}(2018){Zucker}, {Schlafly}, {Speagle}, {Green},
  {Portillo}, {Finkbeiner}, \& {Goodman}}]{Zucker2018}
{Zucker}, C., {Schlafly}, E.~F., {Speagle}, J.~S., {et~al.} 2018,
  \href{http://dx.doi.org/10.3847/1538-4357/aae97c}{\JournalTitle{\apj}, 869,
  83}

\end{thebibliography}

\end{document}